\documentclass[a4paper,11pt]{article}
\pdfoutput=1
\usepackage{amsmath,amssymb,xcolor}
\usepackage[utf8x]{inputenc}
\usepackage{graphicx,subfigure}
\usepackage[colorlinks=true]{hyperref}
\graphicspath{{figures/}{./}}
\usepackage{multirow}
\newcommand{\ok}{$\checkmark$}
\newcommand{\no}{$\times$}

\title{Models with radiative neutrino masses and viable dark matter candidates}
\author{
 Diego Restrepo\footnote{restrepo@udea.edu.co},~Oscar Zapata\footnote{ozapata@fisica.udea.edu.co},\\
\it \small  Instituto de F\'{i}sica, Universidad de Antioquia,\\
\it \small  Calle 70 No. 52-21, Medell\'{i}n, Colombia\\[4mm]
Carlos E. Yaguna\footnote{carlos.yaguna@uni-muenster.de} \\ 
\it \small Institut f\"ur Theoretische Physik, Universit\"at M\"unster,\\
\it \small Wilhelm-Klemm-Stra\ss e 9, D-48149 M\"unster, Germany}
\begin{document}
\maketitle
\begin{abstract}
  We provide a list of particle physics models at the TeV-scale that
  are compatible with neutrino masses and dark matter.
  In these models, the Standard Model particle content is extended
  with a small number ($\leq 4$) of scalar and fermion fields
  transforming as singlets, doublets or triplets under $SU(2)$, and
  neutrino masses are generated radiatively via 1-loop diagrams.
  The dark matter candidates are stabilized by a $Z_2$ symmetry and
  are in general mixtures of the neutral components of such new
  multiplets.
  We describe the particle content of each of these models and
  determine the conditions under which they are consistent with
  current data.
  We find a total of $35$ viable models, most of which have not been
  previously studied in the literature.
  There is a great potential to test these models at the LHC not only
  due to the TeV-scale masses of the new fields but also because about
  half of the viable models contain particles with exotic electric
  charges, which give rise to background-free signals.
  Our results should serve as a first step for detailed analysis of
  models that can simultaneously account for dark matter and neutrino
  masses.

\end{abstract}
\section{Introduction}
The evidence for physics beyond the Standard Model rests on two
pillars: neutrino masses and dark matter, both solidly supported by
the experimental data.
Cosmological observations show that about $25\%$ of the energy-density
of the Universe consists of a new form of matter usually called
\emph{dark matter} \cite{Ade:2013zuv,Komatsu:2010fb}.
The elementary particle responsible for it should be neutral and
stable and, to be consistent with structure formation, it must behave
as \emph{cold} dark matter.
Because none of the known elementary particles satisfy these
conditions, a new particle, the so-called dark matter candidate, is
required.
We still do not know, however, what this new particle is (its mass,
spin, interactions, etc.) or what is the new physics scenario in which
it appears.  Many possibilities have been considered throughout the
years (supersymmetry, extra-dimensions, minimal models, etc.) but the
current experimental data does not point toward any specific solution.

The observation of neutrino oscillations
\cite{Fukuda:1998mi,Ahmad:2002jz,Araki:2004mb,Adamson:2008zt} implies
that neutrinos have non-zero masses --a fact that cannot explained
within the SM.
In extensions of the SM, neutrino masses can be generated in several
ways: the seesaw mechanism \cite{Minkowski:1977sc,Yanagida:1979as,GellMann:1980vs,Mohapatra:1979ia,Magg:1980ut,Schechter:1980gr,Wetterich:1981bx,Lazarides:1980nt,Mohapatra:1980yp,Cheng:1980qt,Foot:1988aq} the inverse-seesaw \cite{Wyler:1982dd,Mohapatra:1986bd,Ma:1987zm,Dev:2012sg}, via loops \cite{Zee:1980ai,Babu:1988ki,Ma:1998dn,Pilaftsis:1991ug}, R-parity
violation~\cite{Hirsch:2000ef}, flavor symmetries \cite{Altarelli:2010gt}, etc.
Some of them require new physics at very large scales whereas others
can be realized at the electroweak scale.
The current experimental data is however inconclusive and does not
tell us what is the actual mechanism that gives rise to neutrino
masses.
It is clear, though, that massive neutrinos and dark matter are both
part of nature and should be incorporated in models of physics beyond
the Standard Model.
In other words, it makes sense to require that models of new physics
be able to explain the dark matter density of the Universe and to
account for the observed pattern of neutrino masses and mixing angles.
Normally, dark matter and neutrino masses are treated as separate
problems, with many models trying to explain one or the other.
It may be, though, that they are related to each other and that, in
addition, both originate from new physics at the TeV scale --a scale
that can be tested at the LHC.

A simple example of this idea is the model proposed by Ma in
\cite{Ma:2006km}~\footnote{There are some earlier works which relate
  neutrino masses generated at tree
  loops~\cite{Krauss:2002px,Cheung:2004xm} with dark matter.}.  In
it, the SM is extended with three singlet fermions and an additional
scalar doublet, all of them odd under a discrete $Z_2$ symmetry.
Neutrino masses are then generated at one loop and the lightest odd
particle, either a fermion or a scalar, is stable and, if neutral, a
good dark matter candidate.  As we will show, this model is only one
example of a very large family of models which can simultaneously
account for neutrino masses and the dark matter.  The common feature
to all these models is that neutrino masses arise radiatively at the
1-loop level.

Recently, a systematic study of all the possible realizations of the Weinberg operator for neutrino masses at 1-loop was presented \cite{Bonnet:2012kz}.
They considered new fermion and scalar fields transforming as singlet, doublets or triplets of $SU(2)$ and with arbitrary values of the hypercharge.
Their main result is the complete list of sets of fields which give rise to neutrino masses at 1-loop --classified according to the topology of the diagram.
We will use such  list as our starting point and study which of those sets can account also for the dark matter and under what conditions.
Specifically, we look for models containing a stable and neutral particle that are compatible with current bounds from accelerator and dark matter experiments, and with the observed value of the dark matter density.  To ensure the stability of the dark matter particle a $Z_2$ symmetry will be imposed.
Under it, the new fields are odd while the SM particles are even.   Requiring that at least one of the new multiplets contains a neutral particle (the dark  matter candidate) constraints the hypercharge to a few different values  within a given set.
Each of those values of the hypercharge will define, therefore, a different model, with a given particle content. Our results show that there exists $35$ non-equivalent models that can simultaneously account for dark matter and neutrino masses, and that most of them have not been previously studied  in the literature.

The rest of the paper is organized as follows.
In the next section, we introduce our notation and explain the procedure we follow to analyze the dark matter compatibility of the different models.
Our main results are presented in Sections \ref{sec:t1-1} to \ref{sec:t4}, where we describe in detail each of the viable models consistent with dark matter and neutrino masses, classified according to the topology of the 1-loop diagram that gives rise to no-zero $\nu$ masses.
At the end of each of those sections,  our findings are summarized in a simple table.  In Section \ref{sec:disc} we briefly discuss possible extensions of our results and in section \ref{sec:con} we present our conclusions. For reference, in the appendices the viable models are categorized by the number of fields, and the complete list of models that are contained within others is provided.

\section{Preliminaries}
\label{sec:pre}
Throughout this paper, we will closely follow the notation from \cite{Bonnet:2012kz} regarding topologies, fields, and models.
Thus, the different topologies will be denoted  by T1-1, T-3, T-4-2-i, etc. according to their figure 3, and the new  fields are denoted as $X_Y^\mathcal{L}$, where $X$ refers to its $SU(2)$ transformation ($1$ for singlet, $2$ for doublet, $3$ for triplet), $\mathcal{L}$ refers to its Lorentz nature ($S$ for scalar, $F$ for fermion), and $Y\equiv 2 (Q-I_3)$ denotes its hypercharge.  The models that are compatible with neutrino masses are denoted  according to the topology and to the order in which they appear in their Tables $2$-$5$.
We see, for example, that according to Table 2 \cite{Bonnet:2012kz} there are $8$ different sets associated to the topology T1-1.
We will denote them, from top to bottom, by T1-1-A to T1-1-H --and so on for all other topologies.  The parameter $\alpha$ appearing in these tables is an integer that  determines the hypercharges of the different fields. 

Our starting point is then the list of possible field assignments that are compatible with neutrino masses classified according to the topology of the 1-loop diagram, as it appears in Tables 2-5 \cite{Bonnet:2012kz}.
It must be emphasized that such list is exhaustive --it includes all the possible realizations of the Weinberg operator at 1-loop with additional fields transforming as singlets, doublets, or triplets of $SU(2)$\footnote{There are additional possibilities if  one allows for larger representations, as discussed recently in \cite{Law:2013saa}.}.
Our goal is to determine which of those models can also account for the dark matter and under what conditions. 

A  dark matter candidate should be stable, color and electrically neutral, and must be consistent with current bounds from dark matter experiments, especially direct detection constraints.
We will impose all these conditions to obtain a set of viable models.
We will also highlight those models containing particles with exotic charges, as they can be more easily tested at the LHC. 

To guarantee the stability of the dark matter particle, a discrete symmetry is usually imposed.
This discrete symmetry, typically a $Z_2$, plays the same role as R-parity in supersymmetric models, rendering the lightest odd particle stable.
We will assume a $Z_2$ symmetry under which   the new fields (those mediating the 1-loop diagram for neutrino masses) are odd while the SM fields are even.
This $Z_2$ symmetry imposes, therefore, a first restriction on these models as it prevents us from associating any of the particles in the loop (odd under the $Z_2$) with SM particles (even under the $Z_2$).
The Zee model~\cite{Zee:1980ai,AristizabalSierra:2006ri}, for example, is a radiative model of neutrino masses (it corresponds to the topology T1-2) that cannot accommodate dark matter within this framework because the loop diagram involves the lepton fields.  In certain cases, this $Z_2$ symmetry is also required to prevent a tree-level contribution to neutrino masses.
In our description of the models, we will explicitly mention when that is the case.

The new fields that generate neutrino masses can in principle have a non-zero color charge, as illustrated by the models discussed in \cite{FileviezPerez:2009ud,Choubey:2012ux}.  Since the lepton doublets and the Higgs field are color neutral (the external legs of the Weinberg operator), as soon as one fixes the color index of one of the internal fields all other are also fixed. Given that we want one of these internal fields to contain the  dark matter particle, which should be color neutral, one of them should be a color singlet and, as a result, all of them should be color singlets. This is another condition that the requirement of dark matter imposes on these models. It eliminates the possibility of having colored particles mediating the 1-loop diagrams for neutrino masses.

A more important restriction comes from requiring a neutral particle in the spectrum.  By demanding $Q=T_3+Y/2=0$ such that $Y=-2T_3$ for at least one particle, we can obtain the values of the parameter $\alpha$ that are consistent with dark matter for each field assignment.  Each such value of $\alpha$ determines what we call a model of radiative neutrino masses and dark matter.
In the next sections, we will obtain the full list of such models.

Once we have a model, we want to make sure that it is consistent with present data.
We must ensure, for example, that the spectrum does not contain new charged stable particles, as they would have been easily observed in accelerators.  For theoretical reasons, we must also enforce the requirement of anomaly cancellation, which  translates into  fermions that must be vector-like.
Dark matter constraints are also important.
The dark matter candidates  in these models can be scalars or fermions and in general they are mixtures of $SU(2)$ singlet, doublet, and triplet representations.
We must ensure that they can account for the observed dark matter density as measured by the WMAP \cite{Komatsu:2010fb} and PLANCK \cite{Ade:2013zuv} satellites.
As usual, we assume the dark matter  density to be the result of a freeze-out process in the early Universe.  In that case, it is already known that the dark matter candidates we consider yield the observed relic density for masses in the TeV range \cite{Cirelli:2005uq} --the precise value depending on the details of the specific model under consideration.
Since it is not our goal to determine the regions that are consistent with the dark matter constraint for each model, we will not strictly enforce the bound on the relic density.
We simply keep in mind that it is always possible to find regions in the parameter space of these models where it can be satisfied and that those regions necessarily feature dark matter masses around the TeV scale.  
Dark matter direct detection experiments will play a crucial role in our analysis as they exclude dark matter candidates that can elastically scatter on nuclei via  $Z$-mediated diagrams.
Specifically they allow to exclude  as dark matter candidates doublet or triplet fermions and triplet scalars, provided they have nonzero hypercharge.
As we will see this is an important restriction that rules many different scenarios out.
Indirect detection experiments, on the other hand, are not expected to be as relevant so we do not impose any restriction based on them.
In section \ref{sec:disc} we comment about their potential.  

One of the advantages of having scenarios where dark matter and neutrino masses are simultaneously addressed is that the masses of the new particles naturally comes out to be, via the freeze-out mechanism in the early Universe, at the TeV scale.
Consequently, these models can be probed in collider searches at the LHC.
With this in mind, we will point out which of the viable models contain particles with exotic electric charges (doubly-charged fermions or scalars), as they give rise to background-free signals which can be more easily searched for in the LHC data.   

We are now ready to present our results: the full set of models with radiative neutrino masses that are compatible with dark matter.
To facilitate the analysis, we will classify the models according to the topology of the 1-loop diagram for neutrino masses, with each topology studied in a different section.
For each field assignment compatible with neutrino masses, we will determine the values of $\alpha$ (the hypercharges) that allow for a dark matter particle, show the particles it contain, find  the dark matter candidates and check whether it is consistent with the bounds previously mentioned.
As we will see, within a  given topology some models are actually equivalent to others --they contain the same particle content.
We will include a remark when that is the case and count only the non-equivalent models.
At the end of each section, a table that summarizes the viable models is provided. 

\section{Models from topology T1-1}
\label{sec:t1-1}

\begin{figure}[th]
\begin{center}
 \includegraphics[scale=1.0]{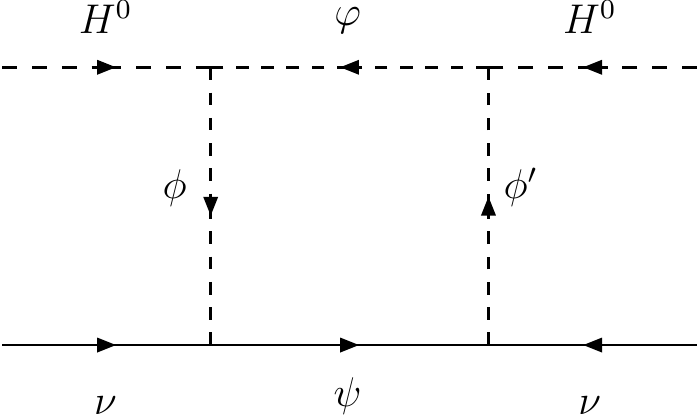}
\caption{One-loop contribution to neutrino mass in the T1-1 models.}
\label{fig:T1-1}
\end{center}
\end{figure}
In this section we consider the models belonging to the topology T1-1.
The 1-loop contribution to the neutrino mass matrix is given in this case by the diagram shown in figure \ref{fig:T1-1}.
All these models include 3 scalars and one fermion field, which are odd under the $Z_2$ symmetry.
Excluding the hypercharge, there are $8$ different field assignments that are consistent with neutrino masses: T1-1-A,...,T1-1-H.
In each of them, we find models containing  a dark matter candidate.
Let us see these viable possibilities one by one.    

\subsection{Model T1-1-A}
\begin{table}[h]
\begin{center}
\begin{tabular}{|c|c|c|c|}
\hline
 $\varphi$ & $\phi'$ & $\psi$ & $\phi$  \\
\hline
$1^{S}_{\alpha}$ &  $2^{S}_{\alpha-1}$ & $1^{F}_{\alpha}$ & $2^{S}_{\alpha+1}$ \\
\hline
\end{tabular}
\end{center}
\caption{\it Model T1-1-A.  }
\label{T1-1-A}
\end{table}

In this kind of model, the fermion is an $SU(2)$ singlet as is one of the scalars.
The other two scalars are $SU(2)$ doublets --see Table \ref{T1-1-A}.
The three values of $\alpha$ that are consistent with dark matter are
\begin{itemize}
\item $\alpha=0:$\hspace{0.5cm} $\varphi^{0}_{0},\hspace{1cm} \phi'_{-1}=(\phi'^{0},\phi'^{-}),\hspace{1cm} \psi^{0}_{0}\hspace{1cm} \phi_{1}=(\phi^{+},\phi^{0})$.\\[2mm]
Since $\phi'=\phi^\dagger$ for this value of $\alpha$ only 3 additional fields are actually required in this case.
This model allows for singlet fermionic dark matter and singlet-doublet scalar DM.
For a realization of this model see e.g. \cite{Farzan:2009ji}.  The $Z_2$ symmetry  also prevents a  Type I seesaw contribution to neutrino masses.
\item $\alpha=2:$\hspace{0.5cm} $\varphi^{+}_{2},\hspace{1cm} \phi'_{1}=(\phi'^{+},\phi'^{0}),\hspace{1cm} \psi^{+}_{2}\hspace{1cm} \phi_{3}=(\phi^{++},\phi^{+})$.\\[2mm]
This model only allows for  doublet scalar DM.
Potential signals at the LHC coming from the doubly-charged scalar particle.
The fermion should be vector-like to ensure the cancellation of anomalies. 
\item $\alpha=-2:$\hspace{0.5cm} $\varphi^{-}_{-2},\hspace{1cm} \phi'_{-3}=(\phi'^{-},\phi'^{--}),\hspace{1cm} \psi^{-}_{-2}\hspace{1cm} \phi_{-1}=(\phi^{0},\phi^{-})$.\\[2mm]
Notice that the particle content in this case is equal, up to charge conjugation, to that obtained for $\alpha=2$.
We say therefore that these two models are equivalent to each other. 
\end{itemize}

\subsection{Model T1-1-B}
\begin{table}[h]
\begin{center}
\begin{tabular}{|c|c|c|c|}
\hline
 $\varphi$ & $\phi'$ & $\psi$ & $\phi$  \\
\hline
$1^{S}_{\alpha}$ &  $2^{S}_{\alpha-1}$ & $3^{F}_{\alpha}$ & $2^{S}_{1+\alpha}$   \\
\hline
\end{tabular}
\end{center}
\caption{\it Model T1-1-B.  }
\label{T1-1-B}
\end{table}
This model differs from the previous one in that the fermion is an $SU(2)$ triplet rather than a singlet, as shown in Table \ref{T1-1-B}.  We again find three values of $\alpha$ that allow to have a dark matter candidate: 
\begin{itemize}
\item $\alpha=0:$\hspace{0.5cm} $\varphi^{0}_{0},\hspace{1cm} \phi'_{-1}=(\phi'^{0},\phi'^{-}),\hspace{1cm}\psi_{0}=(\psi^{+},\psi^{0},\psi^{-})\hspace{1cm} \phi_{1}=(\phi^{+},\phi^{0})$.\\[2mm]
This setup allows for triplet fermion dark matter or  mixed singlet-doublet scalar dark matter.  Notice that the number of additional fields gets reduced to three for this value of $\alpha$ as $\phi^\dagger=\phi'$.
The $Z_2$ symmetry is required also to forbid a Type III seesaw contribution to neutrino masses.
\item $\alpha=2:$\hspace{0.5cm} $\varphi^{+}_{2},\hspace{1cm} \phi'_{1}=(\phi'^{+},\phi'^{0}),\hspace{1cm} \psi_{2}=(\psi^{++},\psi^{+},\psi^{0})\hspace{1cm} \phi_{3}=(\phi^{++},\phi^{+})$.\\[2mm]
This value of $\alpha$ allows for doublet scalar DM.
Triplet fermion DM is excluded by direct detection bounds.
Anomaly cancellation implies that the fermion should be vector-like.
The spectrum contains two particles, one fermion and one scalar, with exotic charges. 
\item $\alpha=-2:$\hspace{0.5cm} $\varphi^{-}_{-2},\hspace{1cm} \phi'_{-3}=(\phi'^{-},\phi'^{--}),\hspace{1cm} \psi_{-2}=(\psi^{0},\psi^{-},\psi^{--})\hspace{1cm} \phi_{-1}=(\phi^{0},\phi^{-})$.\\[2mm]
This is equivalent to the $\alpha=2$ case.
\end{itemize}

\subsection{Model T1-1-C}
\begin{table}[h]
\begin{center}
\begin{tabular}{|c|c|c|c|}
\hline
 $\varphi$ & $\phi'$ & $\psi$ & $\phi$  \\
\hline
$2^{S}_{\alpha}$ &  $1^{S}_{\alpha-1}$ & $2^{F}_{\alpha}$ & $1^{S}_{1+\alpha}$   \\
\hline
\end{tabular}
\end{center}
\caption{\it Model T1-1-C.  }
\label{T1-1-C}
\end{table}
In this model, the fermion and one of the scalars are $SU(2)$ doublets whereas the other two scalars are singlets, as shown in Table \ref{T1-1-C}.
We find two values of $\alpha$ compatible with a dark matter particle: 
\begin{itemize}
\item $\alpha=1:$\hspace{0.5cm} $\varphi_{1}=(\varphi^{+},\varphi^{0}),\hspace{1cm} \phi'^{0}_{0},\hspace{1cm} \psi_{1}=(\psi^{+},\psi^{0}),\hspace{1cm} \phi^{+}_{2}$.\\[2mm]
The dark matter candidate is a singlet-doublet scalar.
Doublet fermion dark matter is excluded by direct detection experiments.
The fermion must be vector-like due to the anomaly constraint.
\item $\alpha=-1:$\hspace{0.5cm} $\varphi_{-1}=(\varphi^{0},\varphi^{-}),\hspace{1cm} \phi'^{-}_{-2},\hspace{1cm} \psi_{-1}=(\psi^{0},\psi^{-}),\hspace{1cm} \phi^{0}_{0}$.\\[2mm]
This is equivalent to the $\alpha=1$ case.
\end{itemize}

\subsection{Model T1-1-D}
\begin{table}[h]
\begin{center}
\begin{tabular}{|c|c|c|c|}
\hline
 $\varphi$ & $\phi'$ & $\psi$ & $\phi$  \\
\hline
$2^{S}_{\alpha}$ &  $1^{S}_{\alpha-1}$ & $2^{F}_{\alpha}$ & $3^{S}_{1+\alpha}$   \\
\hline
\end{tabular}
\end{center}
\caption{\it Model T1-1-D.  }
\label{T1-1-D}
\end{table}
This model differs from the previous one in that one of the scalars ($\phi$) is a triplet rather than a singlet.
The field content is shown in Table \ref{T1-1-D}.
Three different values of $\alpha$ lead to a neutral particle in the spectrum:
\begin{itemize}
\item $\alpha=1:$\hspace{0.5cm} $\varphi_{1}=(\varphi^{+},\varphi^{0}),\hspace{1cm} \phi'^{0}_{0},\hspace{1cm} \psi_{1}=(\psi^{+},\psi^{0}),\hspace{1cm} \phi_{2}=(\phi^{++},\phi^{+},\phi^{0})$.\\[2mm]
The dark matter candidate  in this case is a mixture of the neutral components from the singlet, doublet, and triplet scalars.  Fermionic dark matter is excluded by direct detection bounds.
The $Z_2$ symmetry forbids  a Type II seesaw contribution to neutrino masses.
A doubly-charged scalar particle is present in the spectrum.
The fermion doublet $\psi$ should be vector-like to be consistent with anomaly cancellation.
\item $\alpha=-1:$\hspace{0.5cm} $\varphi_{-1}=(\varphi^{0},\varphi^{-}),\hspace{1cm} \phi'^{-}_{-2},\hspace{1cm} \psi_{-1}=(\psi^{0},\psi^{-}),\hspace{1cm}  \phi_{0}=(\phi^{+},\phi^{0},\phi^{-})$.\\[2mm]
For this value of  $\alpha$ the dark matter candidate is a doublet-triplet scalar particle.
Fermionic dark matter is excluded by direct detection bounds. $\psi$ must be vector-like.
\item $\alpha=-3:$\hspace{0.5cm} $\varphi_{-3}=(\varphi^{-},\varphi^{--}),\hspace{0.5cm} \phi'^{--}_{-4},\hspace{0.5cm} \psi_{-3}=(\psi^{-},\psi^{--}),\hspace{0.5cm}  \phi_{-2}=(\phi^{0},\phi^{-},\phi^{--})$.\\[2mm]
This value of $\alpha$ is not consistent with dark matter because $\phi^0$, the only neutral particle in the spectrum, belongs to a scalar triplet with non-zero hypercharge and is therefore excluded by direct detection searches.
\end{itemize}

\subsection{Model T1-1-E}
\begin{table}[h]
\begin{center}
\begin{tabular}{|c|c|c|c|}
\hline
 $\varphi$ & $\phi'$ & $\psi$ & $\phi$  \\
\hline
$2^{S}_{\alpha}$ &  $3^{S}_{\alpha-1}$ & $2^{F}_{\alpha}$ & $1^{S}_{1+\alpha}$   \\
\hline
\end{tabular}
\end{center}
\caption{\it Model T1-1-E.  }
\label{T1-1-E}
\end{table}
This model can be obtained from  T1-1-D by exchanging the roles of  $\phi'$ and $\phi$, as shown in Table \ref{T1-1-E}.
A neutral particle can be found in the spectrum for three different values of  $\alpha$: 
\begin{itemize}
\item $\alpha=1:$\hspace{0.5cm} $\varphi_{1}=(\varphi^{+},\varphi^{0}),\hspace{1cm} \phi'_{0}=(\phi'^{+},\phi'^{0},\phi'^{-}),\hspace{1cm} \psi_{1}=(\psi^{+},\psi^{0}),\hspace{1cm} \phi^{+}_{2}$.\\[2mm]
This model is equivalent to T1-1-D ($\alpha=-1$).
\item $\alpha=-1:$\hspace{0.5cm} $\varphi_{-1}=(\varphi^{0},\varphi^{-}),\hspace{1cm} \phi'_{-2}=(\phi'^{0},\phi'^{-},\phi'^{--}),\hspace{1cm} \psi_{-1}=(\psi^{0},\psi^{-}),\hspace{1cm} \phi^{0}_{0}$.\\[2mm]
This model is equivalent to T1-1-D ($\alpha=1$).
\item $\alpha=3:$\hspace{0.5cm} $\varphi_{3}=(\varphi^{++},\varphi^{+}),\hspace{1cm} \phi'_{2}=(\phi'^{++},\phi'^{+},\phi'^{0}),\hspace{1cm} \psi_{3}=(\psi^{++},\psi^{+}),\hspace{1cm} \phi^{++}_{4}$.\\[2mm]
This possibility is not consistent with dark matter due to the direct detection bounds. 
\end{itemize}
Notice that T1-1-E does not contribute any new viable and non-equivalent models.

\subsection{Model T1-1-F}
\begin{table}[h]
\begin{center}
\begin{tabular}{|c|c|c|c|}
\hline
 $\varphi$ & $\phi'$ & $\psi$ & $\phi$  \\
\hline
$2^{S}_{\alpha}$ &  $3^{S}_{\alpha-1}$ & $2^{F}_{\alpha}$ & $3^{S}_{1+\alpha}$   \\
\hline
\end{tabular}
\end{center}
\caption{\it Model T1-1-F.  }
\label{T1-1-F}
\end{table}
The particle content of this model consists of two triplet scalars, one fermion doublet and one scalar doublet --see Table \ref{T1-1-F}.
The values of $\alpha$ that could be consistent with dark matter are:
\begin{itemize}
\item $\alpha=1:$\hspace{0.5cm} $\varphi_{1}=(\varphi^{+},\varphi^{0}),\hspace{5mm} \phi'_{0}=(\phi'^{+},\phi'^{0},\phi'^{-}),\hspace{5mm} \psi_{1}=(\psi^{+},\psi^{0}),\hspace{5mm} 
\phi_{2}=(\phi^{++},\phi^{+},\phi^{0})$.\\[2mm]
This case allows for doublet-triplet scalar dark matter.
Fermion dark matter is ruled out by direct detection bounds.
The $Z_2$ symmetry forbids a Type II seesaw contribution to neutrino masses.
Potential signals at the LHC coming from a doubly-charged scalar particle.
The fermion must be vector-like.
\item $\alpha=-1:$\hspace{0.3cm} $\varphi_{-1}=(\varphi^{0},\varphi^{-}),\hspace{4mm} \phi'_{-2}=(\phi'^{0},\phi'^{-},\phi'^{--}),\hspace{4mm} \psi_{-1}=(\psi^{0},\psi^{-}),\hspace{4mm} 
\phi_{0}=(\phi^{+},\phi^{0},\phi^{-})$.\\[2mm]
This setup is equivalent to that for $\alpha=1$.
\item $\alpha=3:$\hspace{0.2cm} $\varphi_{3}=(\varphi^{++},\varphi^{+}),\hspace{3mm} \phi'_{2}=(\phi'^{++},\phi'^{+},\phi'^{0}),\hspace{3mm} \psi_{3}=(\psi^{++},\psi^{+}),\hspace{3mm} 
\phi_{4}=(\phi^{+++},\phi^{++},\phi^{+})$.\\[2mm]
This setup is not consistent with dark matter because  $\phi'$ has non-zero hypercharge.
\item $\alpha\!=\!-3:$ $\varphi_{-3}=(\varphi^{-},\varphi^{--}),\hspace{1mm} \phi'_{-4}=(\phi'^{-},\phi'^{--},\phi'^{---}),\hspace{1mm} \psi_{-3}=(\psi^{-},\psi^{--}),\hspace{1mm} 
\phi_{-2}=(\phi^{0},\phi^{-},\phi^{--})$.\\[2mm]
It is equivalent to the case $\alpha=3$ and therefore inconsistent with dark matter.
\end{itemize}

\subsection{Model T1-1-G}
\begin{table}[h]
\begin{center}
\begin{tabular}{|c|c|c|c|}
\hline
 $\varphi$ & $\phi'$ & $\psi$ & $\phi$  \\
\hline
$3^{S}_{\alpha}$ &  $2^{S}_{\alpha-1}$ & $1^{F}_{\alpha}$ & $2^{S}_{1+\alpha}$   \\
\hline
\end{tabular}
\end{center}
\caption{\it Model T1-1-G.  }
\label{T1-1-G}
\end{table}
As shown in Table \ref{T1-1-G}, this model contains a singlet fermion, two doublet scalars, and one triplet scalar.
Neutral  particles can be obtained for three different values of $\alpha$: 
\begin{itemize}
\item $\alpha=0:$\hspace{0.5cm} $\varphi_{0}=(\varphi^{+},\varphi^{0},\varphi^{-}),\hspace{1cm} \phi'_{-1}=(\phi'^{0},\phi'^{-}),\hspace{1cm} \psi^{0}_{0}\hspace{1cm} \phi_{1}=(\phi^{+},\phi^{0})$.\\[2mm]
It  allows for doublet-triplet scalar dark matter and singlet fermion dark matter.
The $Z_2$ symmetry forbids  a Type I seesaw contribution to neutrino masses.
\item $\alpha=2:$\hspace{0.5cm} $\varphi_{2}=(\varphi^{++},\varphi^{+},\varphi^{0}),\hspace{1cm} \phi'_{1}=(\phi'^{+},\phi'^{0}),\hspace{1cm} \psi^{+}_{2}\hspace{1cm} \phi_{3}=(\phi^{++},\phi^{+})$.\\[2mm]
 This value of $\alpha$ allows for doublet-triplet scalar dark matter.
The $Z_2$ symmetry forbids  a Type II seesaw contribution to neutrino masses.
Potential signals at the LHC coming from two scalar particles with exotic charges. $\psi$ should be vector-like.
\item $\alpha=-2:$\hspace{0.5cm} $\varphi_{-2}=(\varphi^{0},\varphi^{-},\varphi^{--}),\hspace{1cm} \phi'_{-3}=(\phi'^{-},\phi'^{--}),\hspace{1cm} \psi^{-}_{-2}\hspace{1cm} \phi_{-1}=(\phi^{0},\phi^{-})$.\\[2mm]
This setup is equivalent to that for $\alpha=2$.
\end{itemize}

\subsection{Model T1-1-H}
\begin{table}[h]
\begin{center}
\begin{tabular}{|c|c|c|c|}
\hline
 $\varphi$ & $\phi'$ & $\psi$ & $\phi$  \\
\hline
$3^{S}_{\alpha}$ &  $2^{S}_{\alpha-1}$ & $3^{F}_{\alpha}$ & $2^{S}_{1+\alpha}$   \\
\hline
\end{tabular}
\end{center}
\caption{\it Model T1-1-H.  }
\label{T1-1-H}
\end{table}
The last model within this topology, T1-1-H contains a fermion triplet, a scalar triplet, and two scalar doublets.
Three values of $\alpha$ are compatible with dark matter:
\begin{itemize}
\item $\alpha=0:$\hspace{0.5cm} $\varphi_{0}=(\varphi^{+},\varphi^{0},\varphi^{-}),\hfill \phi'_{-1}=(\phi'^{0},\phi'^{-}),\hfill\psi_{0}=(\psi^{+},\psi^{0},\psi^{-}),\hfill
\phi_{1}=(\phi^{+},\phi^{0})$.\\[2mm]
This model allows for doublet-triplet scalar dark matter and triplet fermion dark matter.
The $Z_2$ symmetry forbids a  Type III seesaw contribution to neutrino masses.
\item $\alpha=2:$\hspace{0.4cm} $\varphi_{2}=(\varphi^{++},\varphi^{+},\varphi^{0}),\hspace{4mm} \phi'_{1}=(\phi'^{+},\phi'^{0}),\hspace{4mm} \psi_{2}=(\psi^{++},\psi^{+},\psi^{0}),\hspace{4mm}
\phi_{3}=(\phi^{++},\phi^{+})$.\\[2mm]
This value of $\alpha$ allows for doublet-triplet scalar dark matter.
The $Z_2$ symmetry forbids  a Type II seesaw contribution to neutrino masses.
Potential signals at the LHC coming from a doubly-charged fermion and two doubly-charged scalars.
The fermion triplet should be vector-like to be compatible with anomaly cancellation.
\item $\alpha=\!-2\!:$\hspace{0.1cm} $\varphi_{-2}=(\varphi^{0},\varphi^{-},\varphi^{--}),\hspace{1mm} \phi'_{-3}=(\phi'^{-},\phi'^{--}),\hspace{1mm} \psi_{-2}=(\psi^{0},\psi^{-},\psi^{--}),\hspace{1mm}
\phi_{-1}=(\phi^{0},\phi^{-})$.\\[2mm]
This case is equivalent to that for $\alpha=2$.
\end{itemize}

\subsection{Summary of T1-1}

\begin{table}[h]
\begin{center}
\begin{tabular}{|c|c|c|c|c|c|c|c|}
\hline
\multirow{2}{*}{Model} & \multirow{2}{*}{$\alpha$}& \multicolumn{2}{c|}{Fermionic} & \multicolumn{2}{c|}{Scalar} & \multirow{2}{*}{Exotic charges}& \multirow{2}{*}{\# of N'plets}\\
\cline{3-6}
 & & DM & DD & DM & DD & & \\
\hline
\multirow{2}{*}{T1-1-A} & $\pm2$ & \no & \no & $2_{\pm1}$ & $\ok$ & $\ok$ & 4\\
\cline{2-8}
 & $0$ & $1_{0}$ & \ok & $1_0, 2_{\pm1}$ & \ok & \no & 3\\
\hline
\multirow{2}{*}{T1-1-B} & $\pm2$ & $3_{\pm2}$  & \no & $2_{\pm1}$ & $\ok$ & $\ok$ & 4\\
\cline{2-8}
 & $0$ & $3_{0}$ & \ok & $1_0, 2_{\pm1}$ & \ok & \no & 3\\
\hline
T1-1-C & $\pm1$ & $2_{\pm1}$  & \no & $1_0, 2_{\pm1}$ & $\ok$ & $\ok$ & 4\\
\hline
\multirow{2}{*}{T1-1-D} & $1$ & $2_{1}$  & \no & $1_0, 2_{1}, 3_{2}$ & $\ok$ & $\ok$ & 4\\
\cline{2-8}
 & $-1$ & $2_{-1}$ & \no & $2_{-1}, 3_{0}$ & \ok & \no & 4\\
\hline
T1-1-F & $\pm1$ & $2_{\pm1}$ & \no & $2_{\pm1}, 3_{0}, 3_{\pm2}$ & \ok & \ok & 4\\
\hline
\multirow{2}{*}{T1-1-G} & $\pm2$ & \no  & \no & $2_{\pm1}, 3_{\pm2}$ & \ok & \ok & 4\\
\cline{2-8}
 & $0$ & $1_{0}$ & \ok & $2_{\pm1}, 3_{0}$ & \ok & \no & 3\\
\hline
\multirow{2}{*}{T1-1-H} & $\pm2$ & $3_{\pm2}$  & \no & $2_{\pm1}, 3_{\pm2}$ & \ok & \ok & 4\\
\cline{2-8}
 & $0$ & $3_{0}$ & \ok & $2_{\pm1}, 3_{0}$ & \ok & \no & 3\\
\hline
\end{tabular}
\end{center}
\caption{\it Non-equivalent models belonging to the T1-1 topology that are consistent with dark matter.  }
\label{T1}
\end{table}

Table \ref{T1} summarizes our results regarding the T1-1 topology.
In all, we found 12 non-equivalent models that are consistent with dark matter.
All of them admit scalar dark matter and four allow also for fermionic dark matter.
In seven of these models, the spectrum contains particles with exotic electric charges (doubly charged).  Although the T1-1 topology generally requires four additional fields, in four cases that number is actually reduced to three.
%
%
%
%
%
%
%
%
%
\section{Models from topology T1-2}
\label{sec:t1-2}
\begin{figure}[ht]
\begin{center}
 \includegraphics[scale=1.0]{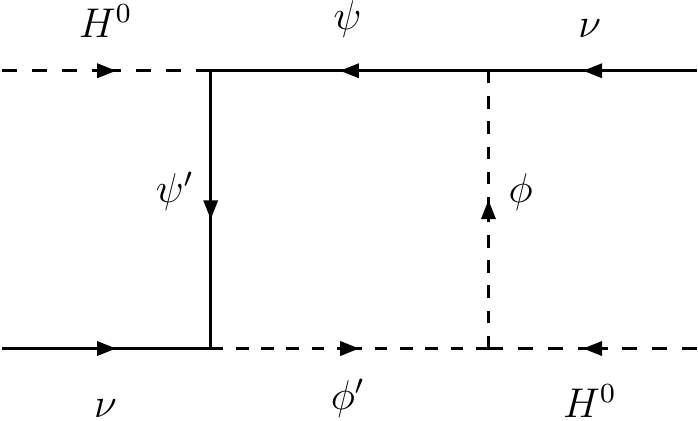}
\caption{One-loop contribution to neutrino mass in the T1-2 models.}
\label{fig:T1-2}
\end{center}
\end{figure}
Models corresponding to the T1-2 topology contain two additional fermions and two new scalars, all assumed to be odd under the $Z_2$ symmetry.
Figure \ref{fig:T1-2} shows the 1-loop contribution to neutrino masses in this topology.
A general discussion about this topology, without establishing the particle content of specific models, was done in~\cite{Farzan:2012ev}.
Eight different field assignments are compatible with neutrino masses: T1-2-A,...,T1-2-H.
As we show next, all of them contain dark matter candidates. 
\subsection{Model T1-2-A}
\begin{table}[h]
\begin{center}
\begin{tabular}{|c|c|c|c|}
\hline
 $\psi$ & $\phi$ & $\phi'$ & $\psi'$  \\
\hline
$1^{F}_{\alpha}$ &  $2^{S}_{1+\alpha}$ & $1^{S}_{\alpha}$ & $2^{F}_{1+\alpha}$   \\
\hline
\end{tabular}
\end{center}
\caption{\it Model T1-2-A.  }
\label{T1-2-A}
\end{table}
This model contains two singlets --one scalar and one fermion-- and two doublets --one scalar and one fermion.
A neutral particle can be found in the spectrum for two different values of $\alpha$:
\begin{itemize}
\item $\alpha=0:$\hspace{0.5cm} $\psi^{0}_{0},\hspace{1cm} \phi_{1}=(\phi^{+},\phi^{0}),\hspace{1cm} \phi'^{0}_{0}\hspace{1cm} \psi'_{1}=(\psi'^{+},\psi'^{0})$.\\[2mm]
This setup allows for singlet-doublet scalar dark matter and singlet-doublet fermionic dark matter.
The $Z_2$ symmetry forbids  a Type I seesaw contribution to neutrino masses. $\psi'$ must be vector-like.\\
\item $\alpha=-2:$\hspace{0.5cm} $\psi^{-}_{-2},\hspace{1cm} \phi_{-1}=(\phi^{0},\phi^{-}),\hspace{1cm} \phi'^{-}_{-2}\hspace{1cm} \psi'_{-1}=(\psi'^{0},\psi'^{-})$.\\[2mm]
In this case only doublet scalar dark matter is allowed.
Fermion doublet dark matter is excluded by direct detection bounds.
Both $\psi$ and $\psi'$ must be vector-like.
\end{itemize}

\subsection{Model T1-2-B}
\begin{table}[h]
\begin{center}
\begin{tabular}{|c|c|c|c|}
\hline
 $\psi$ & $\phi$ & $\phi'$ & $\psi'$  \\
\hline
$1^{F}_{\alpha}$ &  $2^{S}_{1+\alpha}$ & $3^{S}_{\alpha}$ & $2^{F}_{1+\alpha}$   \\
\hline
\end{tabular}
\end{center}
\caption{\it Model T1-2-B.  }
\label{T1-2-B}
\end{table}
This model differs from the previous one in that one of the scalars is a triplet rather than a singlet, as shown in Table \ref{T1-2-B}.
Dark matter candidates are obtained for three values of $\alpha$:
\begin{itemize}
\item $\alpha=0:$\hspace{0.5cm} $\psi^{0}_{0},\hspace{1cm} \phi_{1}=(\phi^{+},\phi^{0}),\hspace{1cm} \phi'_{0}=(\phi'^{+},\phi'^{0},\phi'^{-})\hspace{1cm} \psi'_{1}=(\psi'^{+},\psi'^{0})$.\\[2mm]
This model  admits  doublet-triplet scalar dark matter and singlet-doublet fermionic dark matter.
The $Z_2$ symmetry forbids a Type I seesaw contribution to neutrino masses.
The fermion doublet should be vector-like.
\item $\alpha=-2:$\hspace{0.5cm} $\psi^{-}_{-2},\hspace{1cm} \phi_{-1}=(\phi^{0},\phi^{-}),\hspace{1cm} \phi'_{-2}=(\phi'^{0},\phi'^{-},\phi'^{--})\hspace{1cm} \psi'_{-1}=(\psi'^{0},\psi'^{-})$.\\[2mm]
In this case,  only doublet-triplet scalar dark matter is viable.
The $Z_2$ symmetry forbids a Type II seesaw contributions to neutrino masses.
Notice that the spectrum contains a doubly-charged scalar particle. $\psi$ and $\psi'$ must be vector-like.
\item $\alpha=2:$\hspace{0.5cm} $\psi^{+}_{2},\hspace{1cm} \phi_{3}=(\phi^{++},\phi^{+}),\hspace{1cm} \phi'_{2}=(\phi'^{++},\phi'^{+},\phi'^{0})\hspace{1cm} \psi'_{3}=(\psi'^{++},\psi'^{+})$.\\[2mm]
This possibility is excluded by dark matter direct detection bounds.
\end{itemize}

\subsection{Model T1-2-C}
\begin{table}[h]
\begin{center}
\begin{tabular}{|c|c|c|c|}
\hline
 $\psi$ & $\phi$ & $\phi'$ & $\psi'$  \\
\hline
$2^{F}_{\alpha}$ &  $1^{S}_{1+\alpha}$ & $2^{S}_{\alpha}$ & $1^{F}_{1+\alpha}$   \\
\hline
\end{tabular}
\end{center}
\caption{\it Model T1-2-C.  }
\label{T1-2-C}
\end{table}
The particle content of this model consist of a fermion doublet, a scalar singlet, a scalar doublet and a fermion singlet --see Table \ref{T1-2-C}.
The spectrum contains neutral particles  for two values of $\alpha$:

\begin{itemize}
\item $\alpha=1:$\hspace{0.5cm} $\psi_{1}=(\psi^{+},\psi^{0}),\hspace{1cm} \phi^{+}_{2},\hspace{1cm} \phi'_{1}=(\phi'^{+},\phi'^{0})\hspace{1cm} \psi'^{+}_{2}$.\\[2mm]
This model is equivalent to T1-2-A with $\alpha=-2$.
\item $\alpha=-1:$\hspace{0.5cm} $\psi_{-1}=(\psi^{0},\psi^{-}),\hspace{1cm} \phi^{0}_{0},\hspace{1cm} \phi'_{-1}=(\phi'^{0},\phi'^{-})\hspace{1cm} \psi'^{0}_{0}$.\\[2mm]
This model is equivalent to T1-2-A with $\alpha=0$.
\end{itemize}
Notice that T1-2-C does not give any new non-equivalent models.

\subsection{Model T1-2-D}
\begin{table}[h]
\begin{center}
\begin{tabular}{|c|c|c|c|}
\hline
 $\psi$ & $\phi$ & $\phi'$ & $\psi'$  \\
\hline
$2^{F}_{\alpha}$ &  $1^{S}_{1+\alpha}$ & $2^{S}_{\alpha}$ & $3^{F}_{1+\alpha}$   \\
\hline
\end{tabular}
\end{center}
\caption{\it Model T1-2-D.  }
\label{T1-2-D}
\end{table}
Table \ref{T1-2-D} shows the field content of this model.
It consists of a doublet and a triplet fermion and a singlet and a doublet scalar.
Three values of $\alpha$ might be compatible with dark matter:
\begin{itemize}
\item $\alpha=1:$\hspace{0.5cm} $\psi_{1}=(\psi^{+},\psi^{0}),\hspace{1cm} \phi^{+}_{2},\hspace{1cm} \phi'_{1}=(\phi'^{+},\phi'^{0}),\hspace{1cm} \psi'_{2}=(\psi'^{++},\psi'^{+},\psi'^{0})$.\\[2mm]
Only  doublet scalar dark matter is allowed for this value of $\alpha$.
Fermion dark matter is excluded by direct detection bounds.
The spectrum contains  a doubly-charged fermion.
Both $\psi$ and $\psi'$ must be vector-like to ensure anomaly cancellation.
\item $\alpha=-1:$\hspace{0.5cm} $\psi_{-1}=(\psi^{0},\psi^{-}),\hspace{1cm} \phi^{0}_{0},\hspace{1cm} \phi'_{-1}=(\phi'^{0},\phi'^{-}),\hspace{1cm} \psi'_{0}=(\psi'^{+},\psi'^{0},\psi'^{-})$.\\[2mm]
In this case, the dark matter candidate is a singlet-doublet scalar.
The $Z_2$ symmetry forbids  a Type III seesaw contribution to neutrino masses. $\psi$ must be vector-like. 
\item $\alpha=-3:$\hspace{0.5cm} $\psi_{-3}=(\psi^{-},\psi^{--}),\hspace{7mm} \phi^{-}_{-2},\hspace{7mm} \phi'_{-3}=(\phi'^{-},\phi'^{--}),\hspace{7mm} \psi'_{-2}=(\psi'^{0},\psi'^{-},\psi'^{--})$.\\[2mm]
This case is not consistent with dark matter as the only neutral particle belongs to a fermion triplet with non-zero hypercharge.
\end{itemize}

\subsection{Model T1-2-E}
\begin{table}[h]
\begin{center}
\begin{tabular}{|c|c|c|c|}
\hline
 $\psi$ & $\phi$ & $\phi'$ & $\psi'$  \\
\hline
$2^{F}_{\alpha}$ &  $3^{S}_{1+\alpha}$ & $2^{S}_{\alpha}$ & $1^{F}_{1+\alpha}$   \\
\hline
\end{tabular}
\end{center}
\caption{\it Model T1-2-E.  }
\label{T1-2-E}
\end{table}
This model consists of a singlet and a doublet fermion and a doublet and a triplet scalar, as shown in Table \ref{T1-2-E}.
Possible dark matter candidates are obtained for $\alpha=\pm 1,-3$:
\begin{itemize}
\item $\alpha=1:$\hspace{0.5cm} $\psi_{1}=(\psi^{+},\psi^{0}),\hspace{1cm} \phi_{2}=(\phi^{++},\phi^{+},\phi^{0}),\hspace{1cm} \phi'_{1}=(\phi'^{+},\phi'^{0})\hspace{1cm} \psi'^{+}_{2}$.\\[2mm]
This setup is equivalent to T1-2-B with $\alpha=-2$. 
\item $\alpha=-1:$\hspace{0.5cm} $\psi_{-1}=(\psi^{0},\psi^{-}),\hspace{1cm} \phi_{0}=(\phi^{+},\phi^{0},\phi^{-}),\hspace{1cm} \phi'_{-1}=(\phi'^{0},\phi'^{-})\hspace{1cm} \psi'^{0}_{0}$.\\[2mm]
It is equivalent to T1-2-B with $\alpha=0$.

\item $\alpha=-3:$\hspace{0.5cm} $\psi_{-3}=(\psi^{-},\psi^{--}),\hspace{0.8cm} \phi_{-2}=(\phi^{0},\phi^{-},\phi^{--}),\hspace{0.8cm} \phi'_{-3}=(\phi'^{-},\phi'^{--}),\hspace{0.8cm} \psi'^{-}_{-2}$.\\[2mm]
This case is excluded by direct detection bounds.
\end{itemize}
We do not obtain any viable non-equivalent configurations for this kind of models.

\subsection{Model T1-2-F}
\begin{table}[h]
\begin{center}
\begin{tabular}{|c|c|c|c|}
\hline
 $\psi$ & $\phi$ & $\phi'$ & $\psi'$  \\
\hline
$2^{F}_{\alpha}$ &  $3^{S}_{1+\alpha}$ & $2^{S}_{\alpha}$ & $3^{F}_{1+\alpha}$   \\
\hline
\end{tabular}
\end{center}
\caption{\it Model T1-2-F.  }
\label{T1-2-F}
\end{table}
The field content for this model is illustrated in Table \ref{T1-2-F}.
It consists of two fermions (one doublet and one triplet) and two scalars (one doublet and one triplet).
Dark matter candidates are obtained for $\alpha=\pm 1,-3$:
\begin{itemize}
\item $\alpha=1:$\hspace{0.2cm} $\psi_{1}=(\psi^{+},\psi^{0}),\hspace{5mm} \phi_{2}=(\phi^{++},\phi^{+},\phi^{0}),\hspace{5mm} \phi'_{1}=(\phi'^{+},\phi'^{0})\hspace{5mm} 
\psi'_{2}=(\psi'^{++},\psi'^{+},\psi'^{0})$.\\[2mm]
Only  doublet-triplet scalar dark matter is allowed in this case.
Fermionic dark matter is excluded by direct detection bounds.
The $Z_2$ symmetry forbids  a Type II seesaw contribution to neutrino masses.
There are two particles, one scalar and one fermion, with exotic charges.
Both fermions should be vector-like. 
\item $\alpha=-1:$\hspace{0.2cm} $\psi_{-1}=(\psi^{0},\psi^{-}),\hspace{5mm} \phi_{0}=(\phi^{+},\phi^{0},\phi^{-}),\hspace{5mm} \phi'_{-1}=(\phi'^{0},\phi'^{-})\hspace{5mm}
\psi'_{0}=(\psi'^{+},\psi'^{0},\psi'^{-})$.\\[2mm]
Only  doublet-triplet scalar dark matter is allowed.
The $Z_2$ symmetry forbids  a Type III seesaw contribution to neutrino masses. $\psi$ must be vector-like.
\item $\alpha=-3:\psi_{-3}=(\psi^{-},\psi^{--}),\hspace{1mm} \phi_{-2}=(\phi^{0},\phi^{-},\phi^{--}),\hspace{1mm} \phi'_{-3}=(\phi'^{-},\phi'^{--}),\hspace{1mm}
\psi'_{-2}=(\psi'^{0},\psi'^{-},\psi'^{--})$.\\[2mm]
This case is excluded as both neutral particles belong to triplets with non-zero hypercharge.
\end{itemize}

\subsection{Model T1-2-G}
\begin{table}[h]
\begin{center}
\begin{tabular}{|c|c|c|c|}
\hline
 $\psi$ & $\phi$ & $\phi'$ & $\psi'$  \\
\hline
$3^{F}_{\alpha}$ &  $2^{S}_{1+\alpha}$ & $1^{S}_{\alpha}$ & $2^{F}_{1+\alpha}$   \\
\hline
\end{tabular}
\end{center}
\caption{\it Model T1-2-G.  }
\label{T1-2-G}
\end{table}
This model differs from the previous one in that one of the scalars is a singlet rather than a triplet, as illustrated in Table \ref{T1-2-G}.
Three values of $\alpha$ might be consistent with dark matter:
\begin{itemize}
\item $\alpha=0:$\hspace{0.5cm} $\psi_{0}=(\psi^{+},\psi^{0},\psi^{-}),\hspace{1cm} \phi_{1}=(\phi^{+},\phi^{0}),\hspace{1cm} \phi'^{0}_{0}\hspace{1cm} \psi'_{1}=(\psi'^{+},\psi'^{0})$.\\[2mm]
This model is equivalent to T1-2-D with $\alpha=-1$.
\item $\alpha=-2:$\hspace{0.5cm} $\psi_{-2}=(\psi^{0},\psi^{-},\psi^{--}),\hspace{1cm} \phi_{-1}=(\phi^{0},\phi^{-}),\hspace{1cm} \phi'^{-}_{-2}\hspace{1cm} \psi'_{-1}=(\psi'^{0},\psi'^{-})$.\\[2mm]
This model is equivalent to T1-2-D with $\alpha=1$.
\item $\alpha=2:$\hspace{0.5cm} $\psi_{2}=(\psi^{++},\psi^{+},\psi^{0}),\hspace{1cm} \phi_{3}=(\phi^{++},\phi^{+}),\hspace{1cm} \phi'^{+}_{2}\hspace{1cm} \psi'_{3}=(\psi'^{++},\psi'^{+})$.\\[2mm]
This possibility is excluded by direct detection bounds.
\end{itemize}
Notice that T1-2-G does not offer any new  non-equivalent model.

\subsection{Model T1-2-H}
\begin{table}[h]
\begin{center}
\begin{tabular}{|c|c|c|c|}
\hline
 $\psi$ & $\phi$ & $\phi'$ & $\psi'$  \\
\hline
$3^{F}_{\alpha}$ &  $2^{S}_{1+\alpha}$ & $3^{S}_{\alpha}$ & $2^{F}_{1+\alpha}$   \\
\hline
\end{tabular}
\end{center}
\caption{\it Model T1-2-H.  }
\label{T1-2-H}
\end{table}
This is the last model within this topology.  It consists of two triplets (one scalar and one fermion) and two doublets (one scalar and one fermion), as shown in Table \ref{T1-2-H}.
Dark matter candidates can be obtained for three values of $\alpha$:
\begin{itemize}
\item $\alpha=0:$\hspace{0.5cm} $\psi_{0}=(\psi^{+},\psi^{0},\psi^{-}),\hspace{5mm} \phi_{1}=(\phi^{+},\phi^{0}),\hspace{5mm} \phi'_{0}=(\phi'^{+},\phi'^{0},\phi'^{-}),\hspace{5mm} \psi'_{1}=(\psi'^{+},\psi'^{0})$.\\[2mm]
This model is equivalent to T1-2-F with $\alpha=-1$.
\item $\alpha=\!-2\!:$\hspace{0.1cm} $\psi_{-2}=(\psi^{0},\psi^{-},\psi^{--}),\hspace{2mm} \phi_{-1}=(\phi^{0},\phi^{-}),\hspace{2mm} \phi'_{-2}=(\phi'^{0},\phi'^{-},\phi'^{--}),\hspace{2mm} \psi'_{-1}=(\psi'^{0},\psi'^{-})$.\\[2mm]
This model is equivalent to T1-2-F with $\alpha=1$.
\item $\alpha=2:$\hspace{0.2cm} $\psi_{2}=(\psi^{++},\psi^{+},\psi^{0}),\hspace{3mm} \phi_{3}=(\phi^{++},\phi^{+}),\hspace{3mm} \phi'_{2}=(\phi'^{++},\phi'^{+},\phi'^{0})\hspace{3mm}
 \psi'_{3}=(\psi'^{++},\psi'^{+})$.\\[2mm]
Since both neutral particles belong to triplets with non-zero hypercharge, this possibility is   not consistent with direct detection bounds.
\end{itemize}
No new viable models are obtained from T1-2-H. 

\subsection{Summary of T1-2}
\begin{table}[h]
\begin{center}
\begin{tabular}{|c|c|c|c|c|c|c|c|}
\hline
\multirow{2}{*}{Model} & \multirow{2}{*}{$\alpha$}& \multicolumn{2}{c|}{Fermionic} & \multicolumn{2}{c|}{Scalar} & \multirow{2}{*}{Exotic charges}& \multirow{2}{*}{\# of N'plets}\\
\cline{3-6}
 & & DM & DD & DM & DD & & \\
\hline
\multirow{2}{*}{T1-2-A} & $0$ & $1_0, 2_1$ & $\ok$ & $1_0, 2_{1}$ & \ok & \no & 4\\
\cline{2-8}
 & $-2$ & $2_{-1}$ & \no & $2_{-1}$ & \ok & \no & 4\\
\hline
\multirow{2}{*}{T1-2-B} & $0$ & $1_0, 2_{1}$ & \ok & $2_1, 3_{0}$ & \ok & \no & 4\\
\cline{2-8}
 & $-2$ & $2_{-1}$ & \no & $2_{-1}, 3_{-2}$ & \ok & \ok & 4\\
\hline
\multirow{2}{*}{T1-2-D} & $1$ & $2_{1},3_2$  & \no & $2_{1}$ & $\ok$ & $\ok$ & 4\\
\cline{2-8}
 & $-1$ & $2_{-1}, 3_0$ & \ok & $1_0, 2_{-1}$ & \ok & \no & 4\\
\hline
\multirow{2}{*}{T1-2-F} & $1$ & $2_1, 3_2$  & \no  & $2_1, 3_{2}$ & \ok & \ok & 4\\
\cline{2-8}
 & $-1$ & $2_{-1}, 3_0$ & \ok & $2_{-1}, 3_{0}$ & \ok & \no & 4\\
\hline
\end{tabular}
\end{center}
\caption{\it Non-equivalent models belonging to the T1-2 topology that are consistent with dark matter. }
\label{T1-2}
\end{table}

Our results concerning the T1-2 topology are summarized in Table \ref{T1-2}.  We found 8 non-equivalent models that are consistent with dark matter.
All of them admit scalar dark matter and four allow also for fermionic dark matter.
In three of these models, the spectrum contains particles with exotic electric charges (doubly charged).  In all cases,  four additional fields are required.  

%
%
%
%
%
%
%
%
\section{Models from topology T1-3}
\label{sec:t1-3}
\begin{figure}[ht]
\begin{center}
 \includegraphics[scale=1.0]{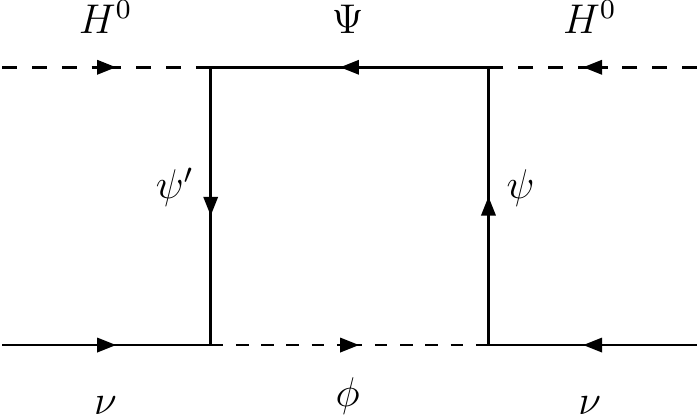}
\caption{One-loop contribution to neutrino mass in the T1-3 models.}
\label{fig:T1-3}
\end{center}
\end{figure}
Models belonging to this topology contain 3 additional fermions ($\psi$, $\psi'$, $\Psi$) and one additional scalar ($\phi$), all odd under the $Z_2$ symmetry. 1-loop neutrino masses are obtained via the diagram in figure \ref{fig:T1-3}.
The eight possible field assignments that are compatible with neutrino masses are denoted as T1-3-A,...,T1-3-H.
Next we examine, for each of them, under what conditions one can obtain viable dark matter candidates.

\subsection{Model T1-3-A}
\begin{table}[h]
\begin{center}
\begin{tabular}{|c|c|c|c|}
\hline
 $\Psi$ & $\psi'$ & $\phi$ & $\psi$  \\
\hline
$1^{F}_{\alpha}$ &  $2^{F}_{1+\alpha}$ & $1^{S}_{\alpha}$ & $2^{F}_{\alpha-1}$   \\
\hline
\end{tabular}
\end{center}
\caption{\it Model T1-3-A.  }
\label{T1-3-A}
\end{table}
In this model, the scalar field and one of the fermions are singlets whereas the other two fermions are  doublets, as shown in Table \ref{T1-3-A}.
The values of $\alpha$ that yield a neutral particle in the spectrum are
\begin{itemize}
\item $\alpha=0:$\hspace{0.5cm} $\Psi^{0}_{0},\hspace{1cm} \psi'_{1}=(\psi'^{+},\psi'^{0}),\hspace{1cm} \phi^{0}_{0},\hspace{1cm} \psi_{-1}=(\psi^{0},\psi^{-})$.\\[2mm]
This model allows for singlet scalar dark matter or singlet-doublet fermion dark matter.
Since $\psi$ and $\psi'$ have opposite hypercharges, the model  is automatically anomaly free.  The $Z_2$ symmetry forbids  a Type I seesaw contribution to neutrino masses.
\item $\alpha=2:$\hspace{0.5cm} $\Psi^{+}_{2},\hspace{1cm} \psi'_{3}=(\psi'^{++},\psi'^{+}),\hspace{1cm} \phi^{+}_{2},\hspace{1cm} \psi_{1}=(\psi^{+},\psi^{0})$.\\[2mm]
Dark matter in this case is excluded by direct detection bounds.
\item $\alpha=-2:$\hspace{0.5cm} $\Psi^{-}_{-2},\hspace{1cm} \psi'_{-1}=(\psi'^{0},\psi'^{-}),\hspace{1cm} \phi^{-}_{-2},\hspace{1cm} \psi_{-3}=(\psi^{-},\psi^{--})$.\\[2mm]
This setup is equivalent to that obtained for $\alpha=2$ and therefore inconsistent with dark matter.

\end{itemize}

\subsection{Model T1-3-B}
\begin{table}[h]
\begin{center}
\begin{tabular}{|c|c|c|c|}
\hline
 $\Psi$ & $\psi'$ & $\phi$ & $\psi$  \\
\hline
$1^{F}_{\alpha}$ &  $2^{F}_{1+\alpha}$ & $3^{S}_{\alpha}$ & $2^{F}_{\alpha-1}$   \\
\hline
\end{tabular}
\end{center}
\caption{\it Model T1-3-B.  }
\label{T1-3-B}
\end{table}
In this model the scalar is a triplet rather than a singlet, as indicated in Table \ref{T1-3-B}.
The spectrum contains a neutral particle for three different values of $\alpha$:
\begin{itemize}
\item $\alpha=0:$\hspace{0.5cm} $\Psi^{0}_{0},\hspace{1cm} \psi'_{1}=(\psi'^{+},\psi'^{0}),\hspace{1cm} \phi_{0}=(\phi^{+},\phi^{0},\phi^{-}),\hspace{1cm} \psi_{-1}=(\psi^{0},\psi^{-})$.\\[2mm]
It allows for triplet scalar dark matter and singlet-doublet fermion dark matter.
The model  is automatically anomaly-free.
The discrete symmetry is required also to prevent a Type I contribution to neutrino masses. 
\item $\alpha=2:$\hspace{0.5cm} $\Psi^{+}_{2},\hspace{1cm} \psi'_{3}=(\psi'^{++},\psi'^{+}),\hspace{1cm} \phi_{2}=(\phi^{++},\phi^{+},\phi^{0}),\hspace{1cm} \psi_{1}=(\psi^{+},\psi^{0})$.\\[2mm]
Both dark matter candidates are excluded by direct detection bounds.
\item $\alpha=-2:$\hspace{0.5cm} $\Psi^{-}_{-2},\hspace{1cm} \psi'_{-1}=(\psi'^{0},\psi'^{-}),\hspace{1cm} \phi_{-2}=(\phi^{0},\phi^{-},\phi^{--}),\hspace{1cm} \psi_{-3}=(\psi^{-},\psi^{--})$.\\[2mm]
Being equivalent to the case $\alpha=2$, this possibility is also inconsistent with dark matter. 
\end{itemize}

\subsection{Model T1-3-C}
\begin{table}[h]
\begin{center}
\begin{tabular}{|c|c|c|c|}
\hline
 $\Psi$ & $\psi'$ & $\phi$ & $\psi$  \\
\hline
$2^{F}_{\alpha}$ &  $1^{F}_{1+\alpha}$& $2^{S}_{\alpha}$ & $1^{F}_{\alpha-1}$   \\
\hline
\end{tabular}
\end{center}
\caption{\it Model T1-3-C.  }
\label{T1-3-C}
\end{table}
In this case, the scalar field is an $SU(2)$ doublet as is one of the fermions.
The other two fermions are $SU(2)$ singlets --see Table \ref{T1-3-C}.
Viable dark matter candidates exist for $\alpha=\pm 1$:
\begin{itemize}
\item $\alpha=1:$\hspace{0.5cm} $\Psi_{1}=(\Psi^{+},\Psi^{0}),\hspace{1cm} \psi'^{+}_{2},\hspace{1cm} \phi_{1}=(\phi^{+},\phi^{0}),\hspace{1cm} \psi^{0}_{0}$.\\[2mm]
This model allows for doublet scalar dark matter and singlet-doublet fermionic dark matter. $\Psi$ and $\psi'$ must be vector-like.
\item $\alpha=-1:$\hspace{0.5cm} $\Psi_{-1}=(\Psi^{0},\Psi^{-}),\hspace{1cm} \psi'^{0}_{0},\hspace{1cm} \phi_{-1}=(\phi^{0},\phi^{-}),\hspace{1cm} \psi^{-}_{-2}$.\\[2mm]
This model is equivalent to that with $\alpha=1$.
\end{itemize}

\subsection{Model T1-3-D}
\begin{table}[h]
\begin{center}
\begin{tabular}{|c|c|c|c|}
\hline
 $\Psi$ & $\psi'$ & $\phi$ & $\psi$  \\
\hline
$2^{F}_{\alpha}$ &  $1^{F}_{1+\alpha}$ & $2^{S}_{\alpha}$ & $3^{F}_{\alpha-1}$   \\
\hline
\end{tabular}
\end{center}
\caption{\it Model T1-3-D.  }
\label{T1-3-D}
\end{table}
Besides a  scalar doublet, this models contains one singlet, one doublet and one triplet fermion, as shown in Table \ref{T1-3-D}.
Dark matter candidates can be found for three different values of $\alpha$:
\begin{itemize}
\item $\alpha=1:$\hspace{0.5cm} $\Psi_{1}=(\Psi^{+},\Psi^{0}),\hspace{1cm} \psi'^{+}_{2},\hspace{1cm} \phi_{1}=(\phi^{+},\phi^{0}),\hspace{1cm} \psi_{0}=(\psi^{+},\psi^{0},\psi^{-})$.\\[2mm]
Fermionic dark matter is excluded by direct detection bounds.
The dark matter particle is the neutral component of the scalar doublet.
The $Z_2$ symmetry forbids  a Type III seesaw contribution to neutrino masses.
All three fermions must be vector-like.

\item $\alpha=-1:$\hspace{0.5cm} $\Psi_{-1}=(\Psi^{0},\Psi^{-}),\hspace{1cm} \psi'^{0}_{0},\hspace{1cm} \phi_{-1}=(\phi^{0},\phi^{-}),\hspace{1cm} \psi_{-2}=(\psi^{0},\psi^{-},\psi^{--})$.\\[2mm]
In this case, the dark matter can be either a  scalar doublet or a   singlet-doublet-triplet fermion.
Potential signals at the LHC coming from a doubly-charged fermion. $\Psi$ and $\psi'$ must be vector-like to be consistent with the cancellation of anomalies.
\item $\alpha=3:$\hspace{0.5cm} $\Psi_{3}=(\Psi^{++},\Psi^{+}),\hspace{1cm} \psi'^{++}_{4},\hspace{1cm} \phi_{3}=(\phi^{++},\phi^{+}),\hspace{1cm} \psi_{2}=(\psi^{++},\psi^{+},\psi^{0})$.\\[2mm]
Dark matter is excluded by direct detection constraints.
\end{itemize}

\subsection{Model T1-3-E}
\begin{table}[h]
\begin{center}
\begin{tabular}{|c|c|c|c|}
\hline
 $\Psi$ & $\psi'$ & $\phi$ & $\psi$  \\
\hline
$2^{F}_{\alpha}$ &  $3^{F}_{1+\alpha}$ & $2^{S}_{\alpha}$ & $1^{F}_{\alpha-1}$   \\
\hline
\end{tabular}
\end{center}
\caption{\it Model T1-3-E.  }
\label{T1-3-E}
\end{table}
The particle content consists of a singlet, a doublet and a triplet fermion, and one scalar doublet.
Dark matter candidates can be obtained for $\alpha=\pm1, 3$:
\begin{itemize}
\item $\alpha=1:$\hspace{0.5cm} $\Psi_{1}=(\Psi^{+},\Psi^{0}),\hspace{1cm} \psi'_{2}=(\psi'^{++},\psi'^{+},\psi'^{0}),\hspace{1cm} \phi_{1}=(\phi^{+},\phi^{0}),\hspace{1cm} \psi^{0}_{0}$.\\[2mm]
This model is equivalent to T1-3-D with $\alpha=-1$.
\item $\alpha=-1:$\hspace{0.5cm} $\Psi_{-1}=(\Psi^{0},\Psi^{-}),\hspace{1cm} \psi'_{0}=(\psi'^{+},\psi'^{0},\psi'^{-}),\hspace{1cm} \phi_{-1}=(\phi^{0},\phi^{-}),\hspace{1cm} \psi^{-}_{-2}$.\\[2mm]
This model is equivalent to T1-3-D with $\alpha=1$.
\item $\alpha=-3:$\hspace{0.5cm} $\Psi_{-3}=(\Psi^{-},\Psi^{--}),\hspace{5mm} \psi'_{-2}=(\psi'^{0},\psi'^{-},\psi'^{--}),\hspace{5mm} \phi_{-3}=(\phi^{-},\phi^{--}),\hspace{5mm} \psi^{--}_{-4}$.\\[2mm]
This possibility is not consistent with dark matter direct detection bounds.
\end{itemize}
Notice that T1-3-E does not yield any new non-equivalent configurations.

\subsection{Model T1-3-F}
\begin{table}[h]
\begin{center}
\begin{tabular}{|c|c|c|c|}
\hline
 $\Psi$ & $\psi'$ & $\phi$ & $\psi$  \\
\hline
$2^{F}_{\alpha}$ &  $3^{F}_{1+\alpha}$ & $2^{S}_{\alpha}$ & $3^{F}_{\alpha-1}$   \\
\hline
\end{tabular}
\end{center}
\caption{\it Model T1-3-F.}
\label{T1-3-F}
\end{table}
Two fermion triplets, one fermion doublet and one scalar doublet are part of this model --see Table \ref{T1-3-F}.
Dark matter candidates are obtained for $\alpha=\pm 1,\pm3$: 
\begin{itemize}
\item $\alpha=1:$\hspace{0.3cm} $\Psi_{1}=(\Psi^{+},\Psi^{0}),\hspace{4mm} \psi'_{2}=(\psi'^{++},\psi'^{+},\psi'^{0}),\hspace{4mm} \phi_{1}=(\phi^{+},\phi^{0}),\hspace{4mm}
\psi_{0} =(\psi^{+},\psi^{0},\psi^{-})$.\\[2mm]
This model allows for doublet scalar dark matter or  doublet-triplet fermionic dark matter.
The $Z_2$ symmetry forbids  a Type III seesaw contribution to neutrino masses.
The spectrum contains a doubly-charged fermion. $\Psi$ and $\psi'$ must be vector-like.
\item $\alpha=-1:$\hspace{0.2cm} $\Psi_{-1}=(\Psi^{0},\Psi^{-}),\hspace{3mm} \psi'_{0}=(\psi'^{+},\psi'^{0},\psi'^{-}),\hspace{3mm} \phi_{-1}=(\phi^{0},\phi^{-}),\hspace{3mm}
\psi_{-2} =(\psi^{0},\psi^{-},\psi^{--})$.\\[2mm]
This setup is equivalent to that with $\alpha=1$.
\item $\alpha=3:$\hspace{0.1cm} $\Psi_{3}=(\Psi^{++},\Psi^{+}),\hspace{2mm} \psi'_{4}=(\psi'^{+++},\psi'^{++},\psi'^{+}),\hspace{2mm} \phi_{3}=(\phi^{++},\phi^{+}),\hspace{2mm}
 \psi_{2}=(\psi^{++},\psi^{+},\psi^{0})$.\\[2mm]
Dark matter is excluded as the neutral particle belongs to a triplet fermion with non-zero hypercharge.
\item $\alpha\!=\!-3:$ $\Psi_{-3}=(\Psi^{-},\Psi^{--}), \psi'_{-2}=(\psi'^{0},\psi'^{-},\psi'^{--}), \phi_{-3}=(\phi^{-},\phi^{--}),\psi_{-4} =(\psi^{-},\psi^{--},\psi^{---})$.\\[2mm]
This possibility,  equivalent to the case $\alpha=3$, is excluded by direct detection bounds. 
\end{itemize}

\subsection{Model T1-3-G}
\begin{table}[h]
\begin{center}
\begin{tabular}{|c|c|c|c|}
\hline
 $\Psi$ & $\psi'$ & $\phi$ & $\psi$  \\
\hline
$3^{F}_{\alpha}$ &  $2^{F}_{1+\alpha}$ & $1^{S}_{\alpha}$ & $2^{F}_{\alpha-1}$   \\
\hline
\end{tabular}
\end{center}
\caption{\it Model T1-3-G.}
\label{T1-3-G}
\end{table}
A slight variation of T1-3-A with a triplet fermion rather than a singlet one --see Table \ref{T1-3-G}.
The spectrum contains neutral particles for three different values of $\alpha$: 
\begin{itemize}
\item $\alpha=0:$\hspace{0.5cm} $\Psi_{0}=(\Psi^{+},\Psi^{0},\Psi^{-}),\hspace{10mm} \psi'_{1}=(\psi'^{+},\psi'^{0}),\hspace{10mm} \phi^{0}_{0},\hspace{10mm} \psi_{-1}=(\psi^{0},\psi^{-})$.\\[2mm]
This model  allows for singlet scalar and doublet-triplet fermion dark matter.  Anomaly cancellation is automatic.
The discrete symmetry prevents a Type III contribution to neutrino masses. 
\item $\alpha=2:$\hspace{0.5cm} $\Psi_{2}=(\Psi^{++},\Psi^{+},\Psi^{0}),\hspace{1cm} \psi'_{3}=(\psi'^{++},\psi'^{+}),\hspace{1cm} \phi^{+}_{2},\hspace{1cm} \psi_{1}=(\psi^{+},\psi^{0})$.\\[2mm]
This setup is excluded by direct detection constraints.
\item $\alpha=-2:$\hspace{0.5cm} $\Psi_{-2}=(\Psi^{0},\Psi^{-},\Psi^{--}),\hspace{0.5cm} \psi'_{-1}=(\psi'^{0},\psi'^{-}),\hspace{0.5cm} \phi^{-}_{-2},\hspace{0.5cm} \psi_{-3}=(\psi^{-},\psi^{--})$.\\[2mm]
It is equivalent to the case $\alpha=2$.
It is not consistent with dark matter.
\end{itemize}

\subsection{Model T1-3-H}
\begin{table}[h]
\begin{center}
\begin{tabular}{|c|c|c|c|}
\hline
 $\Psi$ & $\psi'$ & $\phi$ & $\psi$  \\
\hline
$3^{F}_{\alpha}$ &  $2^{F}_{1+\alpha}$ & $3^{S}_{\alpha}$ & $2^{F}_{\alpha-1}$   \\
\hline
\end{tabular}
\end{center}
\caption{\it Model T1-3-H.  }
\label{T1-3-H}
\end{table}
The last model within the T1-3 topology.
The difference with respect to the previous one is that the scalar is now a triplet rather than a singlet, as shown in Table \ref{T1-3-H}.
Dark matter candidates exists for three values of $\alpha$:
\begin{itemize}
\item $\alpha=0:$\hspace{0.4cm} $\Psi_{0}=(\Psi^{+},\Psi^{0},\Psi^{-}),\hspace{4mm} \psi'_{1}=(\psi'^{+},\psi'^{0}),\hspace{4mm} \phi_{0}=(\phi^{+},\phi^{0},\phi^{-}),\hspace{4mm}\psi_{-1}=(\psi^{0},\psi^{-})$.\\[2mm]
The model  allows for triplet scalar and doublet-triplet fermion DM.
The discrete symmetry prevents a Type III contribution to neutrino masses. 
\item $\alpha=2:$\hspace{0.2cm} $\Psi_{2}=(\Psi^{++},\Psi^{+},\Psi^{0}),\hspace{3mm} \psi'_{3}=(\psi'^{++},\psi'^{+}),\hspace{3mm} \phi_{2}=(\phi^{++},\phi^{+},\phi^{0}),\hspace{3mm}
 \psi_{1}=(\psi^{+},\psi^{0})$.\\[2mm]
Direct detection constraints exclude this possibility.
\item $\alpha\!=\!-2:$\hspace{0.1cm} $\Psi_{-2}=(\Psi^{0},\Psi^{-},\Psi^{--}),\hspace{1mm} \psi'_{-1}=(\psi'^{0},\psi'^{-}),\hspace{1mm} \phi_{-2}=(\phi^{0},\phi^{-},\phi^{--}),\hspace{1mm}
 \psi_{-3}=(\psi^{-},\psi^{--})$.\\[2mm]
Being equivalent to the $\alpha=2$ case, it is also excluded by direct detection.
\end{itemize}

\subsection{Summary of T1-3}

\begin{table}[h]
\begin{center}
\begin{tabular}{|c|c|c|c|c|c|c|c|}
\hline
\multirow{2}{*}{Model} & \multirow{2}{*}{$\alpha$}& \multicolumn{2}{c|}{Fermionic} & \multicolumn{2}{c|}{Scalar} & \multirow{2}{*}{Exotic charges}& \multirow{2}{*}{\# of N'plets}\\
\cline{3-6}
 & & DM & DD & DM & DD & & \\
\hline
T1-3-A & $0$ & $1_0, 2_{\pm1}$ & \ok & $1_{0}$ & \ok & \no & 3\\
\hline
T1-3-B & $0$ & $1_0, 2_{\pm1}$ & \ok & $3_{0}$ & \ok & \no & 3\\
\hline
T1-3-C & $\pm1$ & $1_0, 2_{\pm1}$  & \ok & $2_{1}$ & \ok & \no & 4\\
\hline
\multirow{2}{*}{T1-3-D} & $1$ & $2_{1}, 3_0$ & \ok & $2_{1}$ & \ok & \no & 4\\
\cline{2-8}
 & $-1$ & $1_0, 2_{-1}, 3_{-2}$ & \ok & $2_{-1}$ & \ok & \ok & 4\\
\hline
T1-3-F & $\pm1$ & $2_{\pm1}, 3_0, 3_{\pm2}$ & \ok & $2_{\pm1}$ & \ok & \ok & 4\\
\hline
T1-3-G & $0$ & $2_{\pm1}, 3_0$ & \ok & $1_0$ & \ok & \no & 3\\
\hline
T1-3-H & $0$ & $2_{\pm1}, 3_{0}$ & \ok & $3_{0}$ & \ok & \no & 3\\
\hline
\end{tabular}
\end{center}
\caption{\it Non-equivalent models belonging to the T1-3 topology that are consistent with dark matter.}
\label{T1-3}
\end{table}
Within this topology, we found eight non-equivalent models that are consistent with dark matter, as shown in Table \ref{T1-3}.  Interestingly, all of them admit scalar and fermionic dark matter.   In two of these models, the spectrum contains particles with exotic electric charges (doubly charged).  Although the T1-3 topology generally requires four additional fields, in four cases that number is actually reduced to three.
%
%
%
%
%
%
%
%
%
%

\section{Models from topology T3}
\label{sec:t3}
\begin{figure}[ht]
\begin{center}
 \includegraphics[scale=1.0]{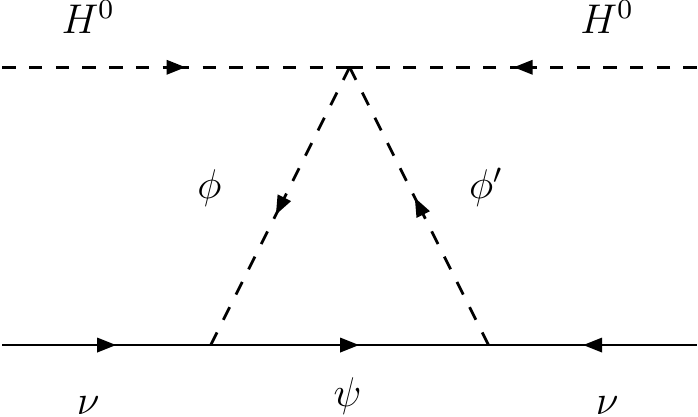}
\caption{One-loop contribution to neutrino mass in the T3 models.}
\label{fig:T3}
\end{center}
\end{figure}
Models from the T3 topology contain only 3 new multiplets, two scalars and one fermion, all odd under the $Z_2$ symmetry.
Figure \ref{fig:T3} shows the diagram that gives rise to neutrino masses in this case.
This topology is the best studied one. It includes the radiative seesaw \cite{Ma:2006km}, its variant with a triplet fermion \cite{Ma:2008cu}, the so-called AMEND \cite{Farzan:2010mr},  and some other possibilities recently discussed in \cite{Law:2013saa}.
Five different field assignments are consistent with non-zero neutrino masses: T3-A, ..., T3-E.
All of them contain viable dark matter candidates.  
\subsection{Model T3-A}
\begin{table}[h]
\begin{center}
\begin{tabular}{|c|c|c|}
\hline
 $\phi'$ & $\phi$ & $\psi$  \\
\hline
$1^{S}_{\alpha}$ & $3^{S}_{2+\alpha}$ & $2^{F}_{1+\alpha}$   \\
\hline
\end{tabular}
\end{center}
\caption{\it Model T3-A.  }
\label{T3-A}
\end{table}
This model consists of a singlet and a triplet scalar, and a fermion doublet --see Table \ref{T3-A}.
The spectrum contains a neutral particle for three different values of $\alpha$:
\begin{itemize}
\item $\alpha=0:$\hspace{0.5cm} $\phi'^{0}_{0},\hspace{1cm} \phi_{2}=(\phi^{++},\phi^{+},\phi^{0}),\hspace{1cm} \psi_{1}=(\psi^{+},\psi^{0})$.\\[2mm]
It allows for singlet-triplet scalar dark matter. 
Fermionic dark matter is excluded by direct detection bounds. 
The $Z_2$ symmetry forbids  a Type II seesaw contribution to neutrino masses. $\psi$ must be vector-like. 
Notice that the spectrum contains one particle with exotic charge: $\phi^{++}$.  
The AMEND model \cite{Farzan:2010mr} is a realization of this setup. It was also analyzed in \cite{Law:2013saa}.
\item $\alpha=-2$: \hspace{0.5cm} $\phi'^{-}_{-2},\hspace{1cm} \phi_{0}=(\phi^{+},\phi^{0},\phi^{-}),\hspace{1cm} \psi_{-1}=(\psi^{0},\psi^{-})$.\\[2mm]
This value of $\alpha$ allows for triplet scalar dark matter.  $\psi$ must be vector-like. This model was discussed in \cite{Law:2013saa}.
\item $\alpha=-4$: \hspace{0.5cm} $\phi'^{--}_{-4},\hspace{1cm} \phi_{-2}=(\phi^{0},\phi^{-},\phi^{--}),\hspace{1cm} \psi_{-3}=(\psi^{-},\psi^{--})$.\\[2mm]
This case is not consistent with direct detection bounds (see \cite{Law:2013saa} for more details about this model).
\end{itemize}

\subsection{Model T3-B}
\begin{table}[h]
\begin{center}
\begin{tabular}{|c|c|c|}
\hline
 $\phi'$ & $\phi$ & $\psi$  \\
\hline
$2^{S}_{\alpha}$ & $2^{S}_{2+\alpha}$ & $1^{F}_{1+\alpha}$   \\
\hline
\end{tabular}
\end{center}
\caption{\it Model T3-B.  }
\label{T3-B}
\end{table}
In this case, the fermion is an $SU(2)$ singlet whereas the scalars are $SU(2)$ doublets --see Table \ref{T3-B}.
Compatibility with dark matter can be obtained for $\alpha=\pm 1,-3$:
\begin{itemize}
\item $\alpha=-1$: \hspace{0.5cm} $\phi'_{-1}=(\phi'^{0},\phi'^{-}),\hspace{1cm} \phi_{1}=(\phi^{+},\phi^{0}),\hspace{1cm} \psi^{0}_{0}$.\\[2mm]
It allows for scalar (doublet) and fermionic (singlet) dark matter. 
Notice that in general this topology requires 3 additional fields but for $\alpha=-1$ (and only for that value) it happens that $\phi^\dagger=\phi'$. 
So at the end only two different additional  fields are required.  This is the well-known radiative seesaw model \cite{Ma:2006km}. 
In this case the $Z_2$ symmetry is also required to prevent a tree-level contribution (Type I seesaw) to neutrino masses.
\item $\alpha=1:$\hspace{0.5cm} $\phi'_{1}=(\phi'^{+},\phi'^{0}),\hspace{1cm} \phi_{3}=(\phi^{++},\phi^{+}),\hspace{1cm} \psi^{+}_{2}$.\\[2mm]
This model allows for scalar doublet dark matter. 
The fermion should be vector-like. 
Potential signals at the LHC coming from particles with exotic charges. 
This model has been discussed in \cite{Aoki:2011yk}. 
\item $\alpha=-3$: \hspace{0.5cm} $\phi'_{-3}=(\phi'^{-},\phi'^{--}),\hspace{1cm} \phi_{-1}=(\phi^{0},\phi^{-}),\hspace{1cm} \psi_{-2}^{-}$.\\[2mm]
It is equivalent to the case $\alpha=1$.
\end{itemize}

\subsection{Model T3-C}
\begin{table}[h]
\begin{center}
\begin{tabular}{|c|c|c|}
\hline
 $\phi'$ & $\phi$ & $\psi$  \\
\hline
$2^{S}_{\alpha}$ & $2^{S}_{2+\alpha}$ & $3^{F}_{1+\alpha}$   \\
\hline
\end{tabular}
\end{center}
\caption{\it Model T3-C.  }
\label{T3-C}
\end{table}
In this case, the fermion is a triplet rather than a singlet (see Table \ref{T3-C}).
Compatibility with dark matter can be obtained for $\alpha=\pm 1,-3$: 
\begin{itemize}
\item $\alpha=-1$: \hspace{0.5cm} $\phi'_{-1}=(\phi'^{0},\phi'^{-}),\hspace{1cm} \phi_{1}=(\phi^{+},\phi^{0}),\hspace{1cm} \psi_{0}=(\psi^{+},\psi^{0},\psi^{-})$.\\[2mm]
This value of $\alpha$  allows for scalar (doublet) and fermionic (triplet) dark matter. 
Notice that  only two additional  fields are needed in this case since $\phi^\dagger=\phi'$. 
This model is known as the radiative seesaw with triplet fermion \cite{Ma:2008cu}. 
The $Z_2$ symmetry is also required to prevent a tree-level contribution (Type III seesaw) to neutrino masses. 
\item $\alpha=+1:$\hspace{0.5cm} $\phi'_{1}=(\phi'^{+},\phi'^{0}),\hspace{1cm} \phi_{3}=(\phi^{++},\phi^{+}),\hspace{1cm} \psi_{2}=(\psi^{++},\psi^{+},\psi^{0})$.\\[2mm]
The model allows only for scalar doublet dark matter since fermion dark matter is ruled out by direct detection bounds. 
The fermion should be vector-like to be consistent with anomaly-cancellation. 
The spectrum contains two doubly-charged particles, one fermion and one scalar. 
See \cite{Law:2013saa} for more details about this model.  
\item $\alpha=-3$: \hspace{0.5cm} $\phi'_{-3}=(\phi'^{-},\phi'^{--}),\hspace{1cm} \phi_{-1}=(\phi^{0},\phi^{-}),\hspace{1cm} \psi_{-2}=(\psi^{0},\psi^{-},\psi^{--})$.\\[2mm]
It is equivalent to the case $\alpha=1$.
\end{itemize}

\subsection{Model T3-D}
\begin{table}[ht]
\begin{center}
\begin{tabular}{|c|c|c|}
\hline
 $\phi'$ & $\phi$ & $\psi$  \\
\hline
$3^{S}_{\alpha}$ & $1^{S}_{2+\alpha}$ & $2^{F}_{1+\alpha}$   \\
\hline
\end{tabular}
\end{center}
\caption{\it Model T3-D.  }
\label{T3-D}
\end{table}
As shown in Table \ref{T3-D}, this model contains a fermion doublet, a singlet scalar and a triplet scalar.
Dark matter candidates can be found for the following values of $\alpha$:
\begin{itemize}
\item $\alpha=0:$\hspace{0.5cm} $\phi'_{0}=(\phi'^{+},\phi'^{0},\phi'^{-}),\hspace{1cm} \phi^{+}_{2},\hspace{1cm} \psi_{1}=(\psi^{+},\psi^{0})$.\\[2mm]
This model is equivalent to T3-A with $\alpha=-2$.
\item $\alpha=-2:$\hspace{0.5cm} $\phi'_{-2}=(\phi'^{0},\phi'^{-},\phi'^{--}),\hspace{1cm} \phi^{0}_{0},\hspace{1cm} \psi_{-1}=(\psi^{0},\psi^{-})$.\\[2mm]
This model is equivalent  to T3-A with $\alpha$=0. 
\item $\alpha=+2:$\hspace{0.5cm} $\phi'_{2}=(\phi'^{++},\phi'^{+},\phi'^{0}),\hspace{1cm} \phi^{++}_{4},\hspace{1cm} \psi_{3}=(\psi^{++},\psi^{+})$.\\[2mm]
This case is not consistent with direct detection constraints.
\end{itemize}
No new non-equivalent configurations were found within this class of models.

\subsection{Model T3-E}
\begin{table}[ht]
\begin{center}
\begin{tabular}{|c|c|c|}
\hline
 $\phi'$ & $\phi$ & $\psi$  \\
\hline
$3^{S}_{\alpha}$ & $3^{S}_{2+\alpha}$ & $2^{F}_{1+\alpha}$   \\
\hline
\end{tabular}
\end{center}
\caption{\it Model T3-E.  }
\label{T3-E}
\end{table}
Two scalar triplets and one fermion doublet are part of this model --see Table \ref{T3-E}.
The spectrum contains neutral particles for four different values of $\alpha$:
\begin{itemize}
\item $\alpha=0:$\hspace{0.5cm} $\phi'_{0}=(\phi'^{+},\phi'^{0},\phi'^{-}),\hspace{1cm} \phi_{2}=(\phi^{++},\phi^{+},\phi^{0}),\hspace{1cm} \psi_{1}=(\psi^{+},\psi^{0})$.\\[2mm]
The model allows only for scalar triplet dark matter. 
Fermion doublet dark matter is ruled out by direct detection bounds. 
The $Z_2$ symmetry forbids  a Type II seesaw contribution to neutrino masses. 
The spectrum contains a doubly-charged scalar particle. 
This model was analyzed in \cite{Law:2013saa}.
\item $\alpha=-2:$\hspace{0.5cm} $\phi'_{-2}=(\phi'^{0},\phi'^{-},\phi'^{--}),\hspace{1cm} \phi_{0}=(\phi^{+},\phi^{0},\phi^{-}),\hspace{1cm} \psi_{-1}=(\psi^{0},\psi^{-})$.\\[2mm]
This is equivalent to the case $\alpha=0$.
\item $\alpha=+2:$\hspace{0.5cm} $\phi'_{2}=(\phi'^{++},\phi'^{+},\phi'^{0}),\hspace{1cm} \phi_{4}=(\phi^{+++},\phi^{++},\phi^{+}),\hspace{1cm} \psi_{3}=(\psi^{++},\psi^{+})$.\\[2mm]
This value of $\alpha$ is excluded because the dark matter candidate belongs to a triplet scalar with non-zero hypercharge (see \cite{Law:2013saa} for more details about this model).
\item $\alpha=-4:$\hspace{0.5cm} $\phi'_{-4}=(\phi'^{-},\phi'^{--},\phi'^{---}),\hspace{1cm} \phi_{-2}=(\phi^{0},\phi^{-},\phi^{--}),\hspace{1cm} \psi_{-1}=(\psi^{-},\psi^{--})$.\\[2mm]
It is equivalent to the case $\alpha=2$ and therefore inconsistent with dark matter.
\end{itemize}

\subsection{Summary of T3}
\begin{table}[ht]
\begin{center}
\begin{tabular}{|c|c|c|c|c|c|c|c|}
\hline
\multirow{2}{*}{Model} & \multirow{2}{*}{$\alpha$}& \multicolumn{2}{c|}{Fermionic} & \multicolumn{2}{c|}{Scalar} & \multirow{2}{*}{Exotic charges}& \multirow{2}{*}{\# of N'plets}\\
\cline{3-6}
 & & DM & DD & DM & DD & & \\
\hline
\multirow{2}{*}{T3-A} & $0$ & $2_{1}$ & \no & $1_0, 3_2$ & \ok & \ok & 3\\
\cline{2-8}
 & $-2$ & $2_{-1}$ & \no & $3_{0}$ & \ok & \no & 3\\
\hline
\multirow{2}{*}{T3-B} & $1,-3$ & \no  & \no & $2_{\pm1}$ & \ok & \ok & 3\\
\cline{2-8}
 & $-1$ & $1_0$ & \ok & $2_{\pm1}$ & \ok & \no & 2\\
\hline
\multirow{2}{*}{T3-C} & $1,-3$ & $3_{\pm2}$  & \no & $2_{\pm1}$ & \ok & \ok & 3\\
\cline{2-8}
 & $-1$ & $3_0$ & \ok & $2_{\pm1}$ & \ok & \no & 2\\
\hline
T3-E & $0,-2$ & $2_{\pm1}$ & \no & $3_{0}, 3_{\pm2}$ & \ok & \ok & 3\\
\hline
\end{tabular}
\end{center}
\caption{\it Non-equivalent models belonging to the T3 topology that are consistent with dark matter.}
\label{T3}
\end{table}

Within this topology, we found seven non-equivalent models that are consistent with dark matter, as shown in Table \ref{T3}.  All of them admit scalar dark matter but only two are compatible with fermionic dark matter.   In four of these models, the spectrum contains particles with exotic electric charges (doubly charged).  Although the T3 topology generally requires three additional fields, in two cases that number is actually reduced to two.
This is the best known scenario for radiative neutrino masses.

%
%
%
%
%
%
%
%
%
\section{Models from topology T4}
\label{sec:t4}
\begin{figure}
  \centering
  \includegraphics[scale=1.0]{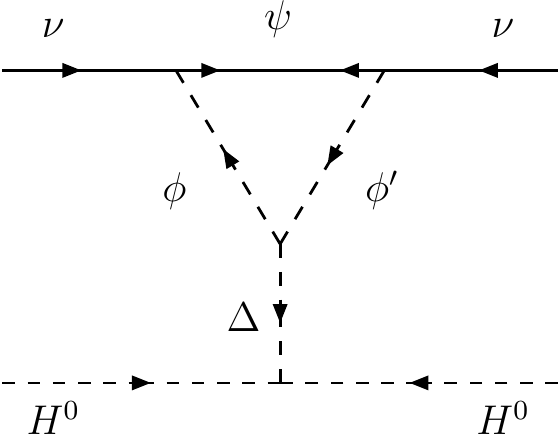}
  \qquad\includegraphics[scale=1.0]{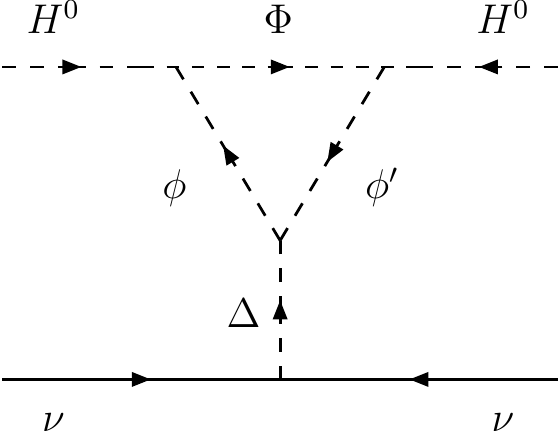}
  \caption{Diagram T4-2-i with Majorana fermions inside the loop, and diagram T4-1-ii.}
  \label{fig:T4-2-i-m}
\end{figure}
One-loop diagrams with T4-like topologies, as the one illustrated in
figure~\ref{fig:T4-2-i-m}, always come accompanied by a tree-level contribution that cannot be forbidden by a discrete or a $U(1)$ symmetry.
As a result, barring an unnatural suppression of  the couplings involved in the tree-level diagrams, the tree-level contribution is generically dominant.
To avoid this,
in \cite{Bonnet:2012kz}, it was proposed to promote the fermions inside the loop to be Majorana fermions, to impose a $Z_2$ symmetry that would forbid a seesaw contribution,  and to assume that all couplings are lepton number conserving.
The first two conditions can be easily satisfied but the last one is not only
ad-hoc but also difficult to implement in a field theory, and should require additional symmetries and particles which are even under $Z_2$,  as in the implementation for T4-1-ii in \cite{Kanemura:2012rj}.
It is clear, in any case, that in the models we consider, where the only additional symmetry is the $Z_2$ required to stabilize the dark matter particle, the tree-level contribution is unavoidable. 
In these models, therefore, neutrino masses do not arise  radiatively  and there is no connection between dark matter and neutrino masses. 
For this reason, we do not consider models from this topology any further.

\section{Discussion}
\label{sec:disc}
In total, we have found $35$ different models of radiative neutrino masses that are compatible with dark matter: $12$ from T1-1, $8$ from T1-2, $8$ from T1-3, and $7$ from T3.
All of them admit scalar dark matter and $18$ allow also for fermionic dark matter.
Particles with exotic electric charges are present in $17$ of these models. Besides the topology, another useful way to classify the viable models is according to the number of different additional fields they require. Obviously, the smaller this number the simpler the model is. As illustrated in the appendix \ref{app:num}, there are only $2$ models with $2$ new fields, $13$ models with $3$ fields and the rest of models contain $4$ additional fields.

As we have seen, the collider and dark matter phenomenology  of many of these viable models have yet to be studied in detail.  In the previous sections, we have only qualitatively described the particle content and the dark matter candidates of each model. A more specific analysis of some of these models is certainly desirable but will be left for future work. Here, we would like to delineate what such a study would entail. 

First of all, one needs to explicitly impose the neutrino mass constraints.
That is, to find the region of the parameter space where the resulting neutrino mass matrix is consistent with the observed pattern of neutrino masses and mixing angles \cite{GonzalezGarcia:2012sz}.  In \cite{Bonnet:2012kz}, the analytic expressions for the neutrino mass matrix can be found  for each topology.
It is important to remark that since the experimental data requires at least two massive neutrinos, more than one generation of a given field is usually required.
The radiative seesaw model (T3-B with $\alpha=-1$), for example, is usually implemented with three generations of the singlet fermion $\psi_0^0$, so as to obtain three light massive neutrinos.
An alternative to multiple generations is to combine different models, as recently done in \cite{Hirsch:2013ola}, where a mixture of T3-B and T3-C was considered.  It must also be kept in mind that more than one topology can contribute to the neutrino mass matrix in a given particle physics model.
The reason for this is that some models are contained within others with more multiplets.
For example, T3-C with $\alpha=-1$ is contained, among others, within T1-1-B with $\alpha=0$ and T1-3-C with $\alpha=\pm1$.
In such cases, both topologies, the one associated with the model itself and the one from the model it contains, contribute to the neutrino mass matrix and must be taken into account when confronting with the experimental data.
To facilitate this task,  in the appendix \ref{sec:within} we provide the full list of models that are contained within others.
The neutrino mass matrix, as expected, depends on the masses and the couplings of the new multiplets.
Rather than verifying whether the neutrino mass matrix is consistent with the data for a given set of masses and couplings, one would like to invert their relation so as to write the couplings in terms of the measured values of the neutrino masses and mixing angles.
That is, to find a parametrization in which the neutrino data is used as an input, in analogy with the Casas-Ibarra parametrization \cite{Casas:2001sr} of the Seesaw mechanism.
That would allow to easily explore and scan the parameter space of the model that is consistent with neutrino masses.

The same loop processes that generate radiative neutrino masses might induce lepton flavor violating processes such as $\mu\to e\gamma$ and $\tau\to\mu\gamma$.
It is important, therefore, to guarantee that their rates are below current experimental bounds \cite{Adam:2013mnn,Aubert:2009ag}.
This constraint is expected to be particularly relevant when the dark matter particle is a singlet fermion, as it  can happen in the radiative seesaw model.
In that case, the dark matter constraint forces the neutrino Yukawa couplings to be large whereas the bounds from $\mu\to  e\gamma$ favored small values \cite{Kubo:2006yx}.
If, on the other hand, the dark matter is an $SU(2)$ doublet or  triplet, the relic density constraint can usually be satisfied via gauge interactions without implying any constraints on the Yukawa couplings that determine the rate of lepton flavor violating processes.   

Regarding dark matter, one certainly needs to obtain the viable parameter space of each model.
That is, to determine the regions that are consistent with the observed value of the dark matter density.
Generic arguments tell us that they do exist and that they correspond to dark matter masses in the TeV range, but one would like to have a more detailed picture.
These regions will strongly depend on the nature of the dark matter particle.
Triplets are expected to be the heaviest followed by doublets and then singlets.
In any case, since in most cases one obtains mixed states, and this mixing depends on the specific parameters of the model,  a case-by-case analysis is necessary.
To make a reliable prediction of the relic density  coannihilation effects must be included and the Sommerfeld enhancement \cite{Hisano:2003ec,Hisano:2004ds,Feng:2010zp} should be taken into account.  

In our analysis, we have included direct detection bounds only partially, simply excluding those dark matter candidates which, having  a direct coupling to the $Z$ boson, feature a spin-independent cross section orders of magnitude larger than current bounds.  It may happen though that certain regions of the parameter space of our viable models  are excluded by the current XENON100 bounds \cite{Aprile:2012nq}.
It is crucial therefore to calculate the detection cross section and to compare it against current data.
In addition, one would like to know what region of the parameter space is going to be probed by future experiments such as XENON1T \cite{Aprile:2012zx}.

We have not taken into account any indirect detection bounds.
They tend to be weaker than direct detection ones and more dependent on astrophysical uncertainties.
Nonetheless, it may happen that due to the Sommerfeld enhancement \cite{Hisano:2003ec,Hisano:2004ds,Feng:2010zp} important bounds can be obtained via the gamma ray signal in some of the viable models we found.
Recent works \cite{Fan:2013faa,Cohen:2013ama}, for example, have shown that a fermion triplet, a so-called wino, can be strongly constrained by indirect detection searches.

Finally, one has to  examine the possibility of excluding/discovering some of these models with LHC data.
Since the new particles do not interact strongly, their production cross sections are not expected to be that large. At the LHC, such cross sections have already been studied in the
literature, as for example in~\cite{Liu:2013gba} ($2^{F,S}_{\pm 1}$ and
$1^{F,S}_{\pm 2}$), \cite{Li:2009mw} ($3^{F}_0$), \cite{FileviezPerez:2008bj}
($3^{S}_0$), \cite{Biondini:2012ny} ($3^{F}_{\pm 2}$ and $2^{F}_{\pm 3}$),
and \cite{Chiang:2012dk} ($3^{S}_{\pm 2}$ and $2^{S}_{\pm 3}$).
At $14\ \text{TeV}$ LHC collisions, the cross section for a $Z_2$-odd
particle of $250\ \text{GeV}$ ranges from a few fb for $1^{S}_{\pm
  2}$, to around $1000\ \text{fb}$ for doubly-charged fermions.
The phenomenology studied there, and the subsequent searches by ATLAS
and CMS of singly~\cite{Salerno:2013ek} and doubly charged
scalars~\cite{Chatrchyan:2012ya,ATLAS:2012hi}, and triplet
fermions~\cite{CMS:2012ra}, have focused in the case in which the new
particles decay into Standard Model fermions. Therefore, these bounds
are not directly applicable to our models.
In fact, as a result of the imposed $Z_2$ symmetry, the lightest
$Z_2$-odd particle (LOP) is stable and all heavier $Z_2$-odd states
decay through others channels into the LOP, with the imprint of large
missing transverse momentum in the detector.
In the inert dark matter models, with only one $Z_2$-odd multiplet, and in
the radiative seesaw models with two of them, the constraints from
direct detection and relic density left little room  for large signal
to background ratios~\cite{Ma:2008cu,Sierra:2008wj,Aoki:2010tf,Klasen:2013jpa}.
In many of the models presented here, specially those with several
DM candidates and extra multiplets no directly related to dark
matter constraints, it could be possible to implement more generic
searches for the heavier than LOP states, like the ones currently
implemented at ATLAS and CMS for electroweak production of
supersymmetric neutralinos, charginos and
sleptons~\cite{ATLAS:2013qla,Chatrchyan:2012pka}. 
In addition, many of our viable models include doubly-charged fermions
and scalars, which should facilitate their study at the LHC. Searches along this line have been already proposed for some
of the models with 3 multiplets in Table~\ref{T3m}: T3-A
($\alpha=0$)~\cite{Farzan:2010mr}, T3-B~\cite{Aoki:2011yk}, and T1-1-1
($\alpha=0$)~\cite{Farzan:2009ji,Farzan:2010fw}.

A possible way to generalize our results is to consider models where the $Z_2$ symmetry we have imposed is,  as
suggested in \cite{Hirsch:2010ru}, the remnant unbroken
subgroup of some flavor symmetry in the lepton sector, which presumably
accounts for neutrino masses and mixings.
This discrete dark matter mechanism, initially applied to $A_4$
in \cite{Hirsch:2010ru,Meloni:2010sk,Boucenna:2011tj}, can be used
with larger non-abelian discrete flavor symmetries
\cite{Meloni:2011cc,Lavoura:2011ry,Boucenna:2012qb}, because they also
have $Z_2$ as a subgroup.
In all these works, the partial spontaneous breaking of the original
flavor symmetry requires the inclusion of at least one even-$Z_2$
scalar multiplet and, as a result, at least one neutrino mass is
generated at tree level, in contrast with the models we have considered. 
It must be kept in mind though that in such flavor models, the odd multiplets under the unbroken $Z_2$ subgroup may induce
one-loop contributions analogous to those  examined in this paper. 
In fact, the radiative seesaw contributions could be implemented in
\cite{Meloni:2010sk} as was already  discussed in
\cite{Lavoura:2011ry}. A more detailed analysis  of this interesting generalization is left for future work.

\section{Conclusions}
\label{sec:con}
Neutrino masses and dark matter provide the only experimental evidences we currently have of physics beyond the Standard Model. 
In this paper, we obtained a comprehensive list of TeV-scale models that can simultaneously explain neutrino masses and account for the dark matter. 
In these models,  non-zero masses are generated radiatively at the 1-loop level and the dark matter candidate is one of the particles mediating the loop diagram --which are assumed to transform as singlets, doublets or triplets of $SU(2)$ and to be odd under a $Z_2$ symmetry. 
In total we found 35 non-equivalent models. 
According to the topology, they are classified in the following way: 12 from T1-1, 8 from T1-2, 8 from T1-3, and 7 from T3. 17 of these viable models admit only scalar dark matter, 18 admit both scalar and fermionic dark matter but none of them is compatible with just fermionic dark matter. 
Half of the models contain new particles with exotic electric charges, making them particularly amenable to searches at the LHC. 
Since most of these models have not been previously studied, they constitute an excellent starting point for future detailed analyses of models of new physics consistent with dark matter and neutrino masses.

\section*{Acknowledgments}
The work of  C.Y. is partially supported by the ``Helmholtz Alliance for Astroparticle Phyics HAP'' funded by the Initiative and Networking Fund of the Helmholtz Association.
D.R. and O.Z. have been partially supported by Sostenibilidad-UdeA and COLCIENCIAS through the grant number 111-556-934918. We want to thank to Diego Aristizabal for helpful discussions. O.Z. is also very grateful to Martin Hirsch for his enlightening discussions.

\bibliographystyle{JHEP}
\bibliography{darkmatter_new}

\appendix
\section{Viable models classified according to the number of multiplets}
\label{app:num}

It is useful to classify the models for neutrino masses and dark matter according to the number of different additional multiplets they require. Tables \ref{T2m}, \ref{T3m} y \ref{T4m} display respectively the non-equivalent models with 2, 3, and 4 multiplets.  In addition, they summarize the main properties of each model.  
\begin{table}[h]
\begin{center}
\begin{tabular}{|c|c|c|c|c|c|c|c|c|}
\hline
\multirow{2}{*}{Model} & \multirow{2}{*}{$\alpha$}& \multicolumn{2}{c|}{Fermionic} & \multicolumn{2}{c|}{Scalar} &Exotic& Additional & \multirow{2}{*}{Viable}\\
\cline{3-6}
 & & DM & DD & DM & DD &charges&fields & \\
\hline
\multirow{1}{*}{T3-B}  & $-1$ & $1_0$ & \ok & $2_{\pm1}$ & \ok & \no & - & \ok \\
\hline
\multirow{1}{*}{T3-C}  & $-1$ & $3_0$ & \ok & $2_{\pm1}$ & \ok & \no & - & \ok\\
\hline
\end{tabular}
\end{center}
\caption{\it Non-equivalent models with 2 multiplets. }
\label{T2m}
\end{table}

\begin{table}[h]
\begin{center}
\begin{tabular}{|c|c|c|c|c|c|c|c|c|}
\hline
\multirow{2}{*}{Model} & \multirow{2}{*}{$\alpha$}& \multicolumn{2}{c|}{Fermionic} & \multicolumn{2}{c|}{Scalar} & Exotic & Additional & \multirow{2}{*}{Viable}\\
\cline{3-6}
 & & DM & DD & DM & DD &  charges & fields &\\
\hline
\multirow{1}{*}{T1-1-A}  & $0$ & $1_{0}$ & \ok & $1_0, 2_{\pm1}$ & \ok & \no & -& \ok\\
\hline
\multirow{1}{*}{T1-1-B} & $0$ & $3_{0}$ & \ok & $1_0, 2_{\pm1}$ & \ok & \no & - & \ok\\
\hline
\multirow{1}{*}{T1-1-G} & $0$ & $1_{0}$ & \ok & $2_{\pm1}, 3_{0}$ & \ok & \no & -  & \ok\\
\hline
\multirow{1}{*}{T1-1-H} & $0$ & $3_{0}$ & \ok & $2_{\pm1}, 3_{0}$ & \ok & \no & -  & \ok\\
\hline
\multirow{1}{*}{T1-3-A}  & $0$ & $1_0, 2_{\pm1}$ & \ok & $1_{0}$ & \ok & \no & - & \ok\\
\hline
\multirow{1}{*}{T1-3-B}  & $0$ & $1_0, 2_{\pm1}$ & \ok & $3_{0}$ & \ok & \no & - & \ok\\
\hline
\multirow{1}{*}{T1-3-G}  & $0$ & $2_{\pm1}, 3_0$ & \ok & $1_0$ & \ok & \no & - & \ok\\
\hline
\multirow{1}{*}{T1-3-H}  & $0$ & $2_{\pm1}, 3_{0}$ & \ok & $3_{0}$ & \ok & \no & - & \ok\\
\hline
\multirow{3}{*}{T3-A} & $0$ & $2_{1}$ & \no & $1_0, 3_2$ & \ok & \ok & - & \ok\\
\cline{2-9}
 & $-2$ & $2_{-1}$ & \no & $3_{0}$ & \ok & \no & $1^S_{-2}$ & \ok \\
\cline{2-9}
 & $-4$ & \no & \no & $3_{-2}$ & \no & \ok & $1^S_{-4},2^F_{-3}$  & \no\\
\hline
\multirow{1}{*}{T3-B} & $1,-3$ & \no  & \no & $2_{\pm1}$ & \ok & \ok & $2^S_{\pm3},1^F_{\pm2}$ & \ok\\
\hline
\multirow{1}{*}{T3-C} & $1,-3$ & $3_{\pm2}$  & \no & $2_{\pm1}$ & \ok & \ok & $2^S_{\pm3}$ & \ok\\
\hline
\multirow{2}{*}{T3-E} & $2,-4$ & \no  & \no & $3_{\pm2}$ & \no & $\ok$ & $3^S_{\pm4},2^F_{\pm3}$ & \no\\
\cline{2-9}
 & $0,-2$ & $2_{\pm1}$ & \no & $3_{0}, 3_{\pm2}$ & \ok & \ok & - & \ok \\
\hline
\end{tabular}
\end{center}
\caption{\it Non-equivalent models with 3 multiplets.  }
\label{T3m}
\end{table}

\begin{table}[h]
\begin{center}
\begin{tabular}{|c|c|c|c|c|c|c|c|c|}
\hline
\multirow{2}{*}{Model} & \multirow{2}{*}{$\alpha$}& \multicolumn{2}{c|}{Fermionic} & \multicolumn{2}{c|}{Scalar} & Exotic& Additional & \multirow{2}{*}{Viable}\\
\cline{3-6}
 & & DM & DD & DM & DD &  charges & Fields &\\
\hline
\multirow{1}{*}{T1-1-A} & $\pm2$ & \no & \no & $2_{\pm1}$ & $\ok$ & $\ok$ &$1^S_{\pm2},2^S_{\pm3},1^F_{\pm2}$  & \ok \\
\hline
\multirow{1}{*}{T1-1-B} & $\pm2$ & $3_{\pm2}$  & \no & $2_{\pm1}$ & $\ok$ & $\ok$ & $1^S_{\pm2},2^S_{\pm3}$ & \ok\\
\hline
T1-1-C & $\pm1$ & $2_{\pm1}$  & \no & $1_0, 2_{\pm1}$ & $\ok$ & $\ok$ & $1^S_{\pm2}$ & \ok\\
\hline
\multirow{3}{*}{T1-1-D} & $1$ & $2_{1}$  & \no & $1_0, 2_{1}, 3_{2}$ & $\ok$ & $\ok$ & - & \ok\\
\cline{2-9}
 & $-1$ & $2_{-1}$ & \no & $2_{-1}, 3_{0}$ & \ok & \no & $1^S_{-2}$ & \ok\\
\cline{2-9}
 & $-3$ & \no & \no & $3_{-2}$ & \no & \ok & $1^S_{-4},2^S_{-3},2^F_{-3}$ & \no\\
\hline
\multirow{2}{*}{T1-1-F} & $\pm3$ & \no  & \no & $3_{\pm2}$ & \no & \ok & $2^S_{\pm3},3^S_{\pm4},2^F_{\pm3}$  & \no\\
\cline{2-9}
 & $\pm1$ & $2_{\pm1}$ & \no & $2_{\pm1}, 3_{0}, 3_{\pm2}$ & \ok & \ok & -  & \ok\\
\hline
\multirow{1}{*}{T1-1-G} & $\pm2$ & \no  & \no & $2_{\pm1}, 3_{\pm2}$ & \ok & \ok & $2^S_{\pm3},1^F_{\pm2}$  & \ok\\
\hline
\multirow{1}{*}{T1-1-H} & $\pm2$ & $3_{\pm2}$  & \no & $2_{\pm1}, 3_{\pm2}$ & \ok & \ok &  $2^S_{\pm3}$   & \ok\\
\hline
\multirow{2}{*}{T1-2-A} & $0$ & $1_0, 2_1$ & $\ok$ & $1_0, 2_{1}$ & \ok & \no & - & \ok\\
\cline{2-9}
 & $-2$ & $2_{-1}$ & \no & $2_{-1}$ & \ok & \no &  $1^S_{-2},1^F_{-2}$  & \ok\\
\hline
\multirow{3}{*}{T1-2-B} & $2$ & \no  & \no & $3_{2}$ & \no & \ok &  $2^S_{3},1^F_{2},2^F_{3}$  & \no\\
\cline{2-9}
 & $0$ & $1_0, 2_{1}$ & \ok & $2_1, 3_{0}$ & \ok & \no & - & \ok \\
\cline{2-9}
 & $-2$ & $2_{-1}$ & \no & $2_{-1}, 3_{-2}$ & \ok & \ok &  $1^F_{-2}$  & \ok\\
\hline
\multirow{3}{*}{T1-2-D} & $1$ & $2_{1},3_2$  & \no & $2_{1}$ & $\ok$ & $\ok$ &  $1^S_{2}$  & \ok \\
\cline{2-9}
 & $-1$ & $2_{-1}, 3_0$ & \ok & $1_0, 2_{-1}$ & \ok & \no & - & \ok \\
\cline{2-9}
 & $-3$ & $3_{-2}$ & \no & \no & \no & \ok &  $1^S_{-2},2^S_{-3},2^F_{-3}$  & \no\\
\hline
\multirow{3}{*}{T1-2-F} & $1$ & $2_1, 3_2$  & \no  & $2_1, 3_{2}$ & \ok & \ok & - & \ok \\
\cline{2-9}
 & $-1$ & $2_{-1}, 3_0$ & \ok & $2_{-1}, 3_{0}$ & \ok & \no & - & \ok \\
\cline{2-9}
 & $-3$ & $3_{-2}$ & \no & $3_{-2}$ & \no & \ok &  $2^S_{-3},2^F_{-3}$  & \no \\
\hline
\multirow{1}{*}{T1-3-A} & $\pm2$ & $2_{\pm1}$ & \no & \no & \no & \ok &  $1^S_{\pm2},1^F_{\pm2},2^F_{\pm3}$  & \no\\
\hline
\multirow{1}{*}{T1-3-B} & $\pm2$ & $2_{\pm1}$  & \no & $3_{\pm2}$ & \no & \ok & $1^F_{\pm2},2^F_{\pm3}$ & \no\\
\hline
T1-3-C & $\pm1$ & $1_0, 2_{\pm1}$  & \ok & $2_{\pm1}$ & \ok & \no & $1^F_{\pm2}$ & \ok\\
\hline
\multirow{3}{*}{T1-3-D} & $3$ & $3_2$  & \no & \no & \no & $\ok$ & $2^S_{3},1^F_{4},2^F_{3}$ & \no\\
\cline{2-9}
 & $1$ & $2_{1}, 3_0$ & \ok & $2_{1}$ & \ok & \no & $1^F_{2}$ & \ok \\
\cline{2-9}
 & $-1$ & $1_0, 2_{-1}, 3_{-2}$ & \ok & $2_{-1}$ & \ok & \ok & - & \ok\\
\hline
\multirow{2}{*}{T1-3-F} & $\pm3$ & $3_{\pm2}$  & \no & \no & \no & \ok & $2^S_{\pm3},2^F_{\pm3},3^F_{\pm4}$ & \no\\
\cline{2-9}
 & $\pm1$ & $2_{\pm1}, 3_0, 3_{\pm2}$ & \ok & $2_{\pm1}$ & \ok & \ok & - & \ok\\
\hline
\multirow{1}{*}{T1-3-G} & $\pm2$ & $2_{\pm1},3_{\pm2}$  & \no  & \no & \no & \ok & $1^S_{\pm2},2^F_{\pm3}$ & \no\\
\hline
\multirow{1}{*}{T1-3-H} & $\pm2$ & $2_{\pm1}, 3_{\pm2}$  & \no & $3_{\pm2}$ & \no & \ok & $2^F_{\pm3}$ & \no\\
\hline
\end{tabular}
\end{center}
\caption{\it Non-equivalent models with 4 multiplets.  }
\label{T4m}
\end{table}
\section{Models contained within others}
\label{sec:within}
When one model is contained within another with more multiplets, the neutrino mass matrix receives two contributions, one associated with the topology of the model itself and another from the model it contains.  Both must be taken into account when confronting the experimental data on neutrino masses and mixing angles.
To facilitate this task,  we provide, in table \ref{tab:cont}, the complete list of viable models that are contained within another. The left (right) panel shows the models with 2 (3) multiplets that are contained in models with 3 (4) multiplets. 
\begin{table}[h]
\begin{minipage}{.5\linewidth}
      \centering
\begin{tabular}{|c|c|c|c|}
\hline
\multirow{2}{*}{Model} & \multirow{2}{*}{$\alpha$}& \multicolumn{2}{c|}{Contained in}\\
\cline{3-4}
 & & Model & $\alpha$ \\
\hline
\multirow{6}{*}{T3-B}  & \multirow{6}{*}{$-1$} & T1-1-A & 0\\
\cline{3-4}
  &  & T1-1-G & 0\\
\cline{3-4}
  &  & T1-2-A & 0\\
\cline{3-4}
  &  & T1-2-B & 0\\
\cline{3-4}
  &  & T1-3-C & $\pm1$\\
\cline{3-4}
  &  & T1-3-D & -1\\
\hline
\multirow{6}{*}{T3-C}  & \multirow{6}{*}{$-1$} & T1-1-B & 0\\
\cline{3-4}
  &  & T1-1-H & 0\\
\cline{3-4}
  &  & T1-2-D & -1\\
\cline{3-4}
  &  & T1-2-F & -1\\
\cline{3-4}
  &  & T1-3-D & 1\\
\cline{3-4}
  &  & T1-3-F & $\pm1$\\
\hline
\end{tabular}
\end{minipage}%
    \begin{minipage}{.5\linewidth}
      \centering
\quad
\begin{tabular}{|c|c|c|c|}
\hline
\multirow{2}{*}{Model} & \multirow{2}{*}{$\alpha$}& \multicolumn{2}{c|}{Contained in}\\
\cline{3-4}
 & & Model & $\alpha$ \\
\hline
T1-1-A  & $0$ & T1-2-A & 0\\
\hline
T1-1-B  & $0$ & T1-2-D & -1\\
\hline
T1-1-G  & $0$ & T1-2-B & 0\\
\hline
T1-1-H  & $0$ & T1-2-F & -1\\
\hline
T1-3-A  & $0$ & T1-2-A & 0\\
\hline
T1-3-B  & $0$ & T1-2-B & 0\\
\hline
T1-3-G  & $0$ & T1-2-D & -1\\
\hline
T1-3-H  & $0$ & T1-2-F & -1\\
\hline
T3-A  & $0$ & T1-1-D & 1\\
\hline
T3-A  & $-2$ & T1-1-D & -1\\
\hline
T3-A  & $-4$ & T1-1-D & -3\\
\hline
T3-B  & $1,-3$ & T1-1-A & $\pm2$\\
\hline
T3-C  & $1,-3$ & T1-1-B & $\pm2$\\
\hline
T3-E  & $2,-4$ & T1-1-F & $\pm3$\\
\hline
T3-E  & $0,-2$ & T1-1-F & $\pm1$\\
\hline
\end{tabular}
\end{minipage} 
\caption{Models with 2 multiplets within models with 3 and 4 multiplets (left) and models with 3 multiplets within models with 4 multiplets (right).}
\label{tab:cont}
\end{table}

\end{document}